\documentclass[twocolumn]{aastex631}

\usepackage{amsmath}    
\usepackage{graphicx}   
\usepackage{hyperref}   
\usepackage{gensymb}    
\usepackage{array}      
\usepackage{booktabs}   
\usepackage{multirow}
\usepackage{float}
\usepackage{footmisc}  

\newcommand{\teff}{$T_{\mathrm{eff}}$}

\newcommand{\numax}{$\nu_{\mathrm{max}}$}

\newcommand{\dnu}{$\Delta\nu$}

\newcommand{\kepler}{\textit{Kepler}}

\newcommand{\totaldet}{290,000}
\newcommand{\bulgedet}{185,000}

\begin{document}

\title{Modeling Asteroseismic Yields for the \textit{Roman} Galactic Bulge Time-Domain Survey}

\author[0009-0008-5554-9144]{Trevor J. Weiss\textsuperscript{\textdagger}}
\affiliation{Department of Physics and Astronomy, California State University, Long Beach, Long Beach, CA 90840, USA}

\author[0009-0006-8874-3846]{Noah J. Downing\textsuperscript{\textdagger}}
\affiliation{Department of Astronomy, The Ohio State University, Columbus, OH 43210, USA}

\author[0000-0002-7549-7766]{Marc H. Pinsonneault}
\affiliation{Department of Astronomy, The Ohio State University, Columbus, OH 43210, USA}

\author[0000-0002-7550-7151]{Joel C. Zinn}
\affiliation{Department of Physics and Astronomy, California State University, Long Beach, Long Beach, CA 90840, USA}

\author[0000-0002-4879-3519]{Dennis Stello}
\affiliation{School of Physics, University of New South Wales, NSW 2052, Australia}

\author[0000-0001-5222-4661]{Timothy R. Bedding}
\affiliation{Sydney Institute for Astronomy (SIfA), School of Physics, University of Sydney, NSW 2006, Australia}

\author[0000-0002-1699-6944]{Kaili Cao}
\affiliation{Center for Cosmology and AstroParticle Physics (CCAPP), The Ohio State University, 191 West Woodruff Ave, Columbus, OH 43210, USA}
\affiliation{Department of Physics, The Ohio State University, 191 West Woodruff Ave, Columbus, OH 43210, USA}

\author[0000-0003-2400-6960]{Marc Hon}
\affiliation{Kavli Institute for Astrophysics and Space Research, Massachusetts Institute of Technology, Cambridge, MA 02139, USA}

\author[0000-0001-9632-2706]{Claudia Reyes}
\affiliation{Research School of Astronomy \& Astrophysics, Australian National University, Canberra ACT 2611, Australia}

\author[0000-0003-0395-9869]{B. Scott Gaudi}
\affiliation{Department of Astronomy, The Ohio State University, Columbus, OH 43210, USA}

\author[0000-0002-4235-6369]{Robert F. Wilson}
\affiliation{Department of Astronomy, University of Maryland, College Park, MD 20742, USA }
\affiliation{NASA Goddard Space Flight Center, Greenbelt, MD 20771, USA}

\author[0000-0001-8832-4488]{Daniel Huber}
\affiliation{Institute for Astronomy, University of Hawai’i, 2680 Woodlawn Drive, Honolulu, HI 96822, USA}
\affiliation{Sydney Institute for Astronomy (SIfA), School of Physics, University of Sydney, NSW 2006, Australia}

\author[0000-0002-0920-809X]{Sanjib Sharma}
\affiliation{Space Telescope Science Institute, 3700 San Martin Drive, Baltimore, MD
21218, USA}

\footnotetext[1]{\textsuperscript{\textdagger} First co-authorship.}

\begin{abstract}
The Galactic Bulge Time-Domain Survey (GBTDS) of the Roman Space Telescope will take high-cadence data of the Galactic bulge. We investigate the asteroseismic potential of this survey for red giants. We simulate the detectability of global asteroseismic frequencies, \numax\ and \dnu, by modifying \textit{Kepler} data to match nominal GBTDS observing strategies, considering different noise models, observing cadences, and detection algorithms. Our baseline case, using conservative assumptions, consistently leads to asteroseismic \numax\ detection probabilities above 80\% for red clump and red giant branch stars brighter than 16th magnitude in Roman's F146 filter. We then inject these detection probabilities into a \textit{Galaxia} model of the bulge to estimate asteroseismic yields. For our nominal case, we detect \totaldet\ stars in total, with \bulgedet\ detections in the bulge. Different assumptions give bulge yields from 135,000 to 349,000 stars. For stars with measured \numax, we find that we can recover \dnu\ in 21\% to 42\% of red clump stars, and 69\% to 92\% of RGB stars. The expected yield and stellar parameter precision we predict for Roman asteroseismology promise to characterize planet-hosting stellar populations and to resolve questions regarding the formation history of the bulge.
\end{abstract}


\section{Introduction} \label{sec:intro}

Time-domain space photometry missions enable a broad range of science, frequently involving topics quite distinct from the main mission goals. The \textit{Kepler} Mission, for example, was designed to study transiting exoplanets \citep{Borucki97, Borucki10}, but has proved extremely valuable for studying stellar oscillations \citep{Gilliland10, Kurtz22}.

The study of stellar oscillations---asteroseismology---can be used to infer stellar mass, radius, and age for large stellar populations \citep{Miglio13, Silva-Aguirre15, Pinsonneault18}. Asteroseismic data can also be used as a training set to infer ages for much larger data sets \citep{Martig16, Ness16b, Mackereth19}. However, these indirect techniques struggle to recover ages for the oldest stars and for those not in the training set \citep{TingRix19, Ciucă21, Leung23}. Therefore, it is highly desirable to obtain more asteroseismic data outside of the solar neighborhood to study stellar populations across the Galaxy.

The Nancy Grace Roman Space Telescope’s Galactic bulge Time-Domain Survey (GBTDS) is one of three Core Community Surveys using the Wide Field Instrument. Its primary purpose is to detect planets through microlensing (\citealt{Penny19}, \citealt{Spergel15}). However, high quality and rapid cadence of the photometry will also enable detections of solar-like oscillations, making Roman uniquely suited to advancing Galactic science. Roman will yield catalogs of asteroseismic stellar parameters for the Galactic bulge, offering insights that will impact a wide range of astrophysical fields (\citealt{Gould15}, \citealt{Huber23}, hereafter G15 and H23, respectively).

The GBTDS will enable the detection of oscillations in red giant branch (RGB) and red clump (RC) stars in the densely populated Galactic bulge, providing crucial insights into the underlying stellar populations. Asteroseismology allows for the precise determinations of mass, radius, and age in evolved stars \citep{Chaplin13, Jackiewicz21}, making it a powerful tool for addressing long-standing questions about the bulge's formation and evolution. In particular, recent studies have suggested the presence of a young stellar population in the bulge \citep[see, e.g.,][]{Bensby17,Joyce23}, a hypothesis that asteroseimic age measurements could directly test.

Red giants oscillate on timescales of hours to days, with amplitudes sufficient for detections at large distances \citep{Miglio21, Hey23}. The central Milky Way is known to contain significant populations of RC stars \citep{Girardi16, Ness16a, Abbott17}, making the bulge an ideal target for asteroseismic studies. In addition to resolving the age distribution of the bulge, asteroseismic constraints on helium abundance, which is otherwise difficult to disentangle from age effects \citep[e.g.,][]{Nataf15}, and radial abundance gradients \citep[e.g.,][]{Hayden15} could refine models of chemical evolution. Moreover, as one of the GBTDS's primary science goals is the detection of microlensed exoplanets, asteroseismology will provide host star ages, offering valuable constraints on planetary evolution models \citep[e.g.][]{Berger20, David21}.

A primary goal of this work is to identify which stellar populations in the bulge are detectable with asteroseismology for use in constructing an asteroseismic target list. Given the $\sim100$ million unique bulge stars expected to be observed with Roman (\citealt{Wislon23}), an asteroseismic target list conditioned on the stellar colors and magnitudes accessible to asteroseismology will be important to reduce the false positive rate and conserve computational resources. As discussed in Section~\ref{sec: conclusion}, science with this sample will also benefit from the stellar population simulations described here in order to define selection/completeness functions for the Roman asteroseismic sample.

The remainder of this section discusses the science and background of using asteroseismology with Roman under differing assumptions for its photometric performance. Section \ref{sec: Simulated Light Curves and Detection Probabilities} describes our methodology for simulating asteroseismic detections. Section \ref{sec: simulated populations} describes our methodology for modeling yields and populations. In Section \ref{sec: Discussion}, we discuss the different yields and the characteristics of our simulated sample. Section \ref{sec: conclusion} summarizes our results and discusses next steps for the project. 

\subsection{\textit{Background on Asteroseismology}}
\label{sec:Background on Asteroseismology}
For solar-like oscillators, turbulence near the stellar surface creates standing wave patterns within the entire star at characteristic frequencies that depend sensitively on mass and radius. When large numbers of stars are involved, it is conventional to use two characteristic frequencies to measure stellar parameters in a process known as ``global asteroseismology". We characterize the observed pattern with a frequency of maximum power, $\nu_{\rm max}$, and the frequency spacing between modes with the same spherical harmonic degree $\ell$, $\Delta\nu$. The former is related to the surface gravity \citep{Brown91, Kjeldsen95, Belkacem11, Hekker20} and the latter is related to the mean density \citep{Ulrich86}; they can therefore be combined to infer mass and radius \citep{Stello08, Kallinger10}:
\begin{equation}
   \frac{R}{R_{\odot}} = \left(f_{\nu_{\rm max}}\frac{\nu_{\rm max}}{\nu_{\rm max,\odot}}\right) \left(f_{\Delta\nu}\frac{\Delta\nu}{\Delta\nu_{\odot}} \right)^{-2} \left( \frac{T_{\rm eff}}{T_{\rm eff,\odot}} \right)^{1/2}
\end{equation}
\begin{equation}\label{eq:2}
   \frac{M}{M_{\odot}} = \left(f_{\nu_{\rm max}}\frac{\nu_{\rm max}}{\nu_{\rm max,\odot}}\right)^3 \left(f_{\Delta\nu}\frac{\Delta\nu}{\Delta\nu_{\odot}} \right)^{-4} \left( \frac{T_{\rm eff}}{T_{\rm eff,\odot}} \right)^{3/2}
\end{equation}
With an independent radius measurement one can instead infer mass using either of the following equations \citep{Ash25}:
\begin{equation}
    \frac{M}{M_{\odot}} = \left(f_{\Delta\nu}\frac{\Delta\nu}{\Delta\nu_{\odot}} \right)^{2} \left( \frac{R}{R_{\odot}} \right)^3
\end{equation}
\begin{equation} \label{eq:4}
    \frac{M}{M_{\odot}} = \left(f_{\nu_{\rm max}}\frac{\nu_{\rm max}}{\nu_{\rm max,\odot}}\right) \left( \frac{T_{\rm eff}}{T_{\rm eff,\odot}} \right)^{1/2} \left( \frac{R}{R_{\odot}} \right)^2
\end{equation}
We include the correction factors $f_{\nu_{\rm max}}$ and $f_{\Delta\nu}$: $f_{\nu_{\rm max}}$ is an empirical correction based on calibration to \textit{Gaia} radii and $f_{\Delta\nu}$ is computed theoretically from stellar models \citep{White11, Sharma16}; both correction factors deviate from unity at the percent level. Note that the correction factors as used here and by \cite{pinsonneault25} are the inverse of the correction factors as defined by \cite{White11, Sharma16, Li-Yaguang23}. To be able to determine stellar age, a mass may be derived from either of the above scaling relations --- in both cases, the mass estimate can be used to perform a stellar age lookup using stellar evolutionary tracks evaluated at a given temperature and metallicity.

Asteroseismology has been revolutionized by space-based time domain missions like CoRoT, \textit{Kepler}/K2, and TESS, which have measured precise masses, radii, and ages for thousands of stars. However, these missions did not observe the bulge. The exception is K2 Campaign 9, but \textit{Kepler}'s large pixels meant the field was too crowded to deliver sufficiently precise light curves for individual stars. There has been work done to study the bulge with ground-based surveys \citep[e.g.,][]{Soszynski13, Hey23}, but it was restricted to the most luminous red giants (the so-called semi-regular variables). We explore here the degree to which the GBTDS will be able to measure the two global asteroseismic frequencies.
\subsection{\textit{Background on Roman/GBTDS}}
\label{sec:Background on Roman}
NASA's next flagship mission, the Nancy Grace Roman Space Telescope, will begin taking science observations in 2027 using two instruments: a wide field imager and a coronagraph. The Wide Field Instrument has an effective field of view of 0.281 deg${^2}$ and a plate scale of 0.11"/pixel. With its Wide Field Instrument, infrared optics and seven filters, Roman will be able to take high-resolution, red-optical to near-IR images with roughly 100 times the FOV of Hubble in one pointing (\citealt{Spergel15}). The GBTDS will have 6 observing seasons of up to $\approx72$ days in length, for a total of $\approx432$ days of observations spread over a 5 year mission. Each field will be observed at approximately 15-minute cadence but the exact value is yet to be decided, so we explore yields given two different cadence scenarios.

The primary goal of the GBTDS is quality imaging and data collection of the Galactic bulge, a population for which it is difficult to estimate precise stellar masses and ages due to high extinction and crowded fields \citep[e.g.,][]{1997ApJ...477..163S}. \textit{Kepler} sampled relatively nearby stars, K2 was restricted to the ecliptic plane, and, while TESS surveys the entire sky, it is much shallower. The bulge will likewise not be accessible to the upcoming PLAnetary Transits and Oscillations of stars (PLATO) mission \citep{2022A&A...658A..31N}. The GBTDS will therefore provide the deepest look into the Galaxy to date and will provide the only survey of the bulge at such high cadence.

\subsection{\textit{Predicting Roman Asteroseismic Yields}}
\label{sec:Predicting Roman Asteroseismic Yields}
This work expands on G15 and H23, which both provided preliminary simulations of asteroseismic detections and population yields for the GBTDS. Such studies are essential for developing potential target lists and understanding the selection functions of the sample, which will be the focus of future work. However, they were both limited in scope. G15 employed a semi-analytic noise model applied to only a few individual stars (rather than a larger, statistically significant sample), demonstrating that asteroseismology may be possible with Roman, but requiring more detailed simulations. H23 expanded on this by exploring cadence variations and more sophisticated models for source counts, but it still used the semi-analytic detection model from G15.

This work expands on G15 and H23 by sampling a wider range of stellar parameter space and considering updated noise models, as described below. In simulating light curves, we used Roman's F146 filter instead of the 2MASS $H$-band approximation used in G15. We utilized dust maps to simulate realistic interstellar dust in the GBTDS fields and generated synthetic stellar populations under various survey strategies. We also reckoned the final detection counts using an SNR-based method, as well as an empirical approach. This provided a holistic sample of expected asteroseismic yields for both \numax\ and \dnu\ given a range of choice for extinction, field selection, noise properties, and detection method, building on and expanding the work of G15 and H23.

\section{Simulated Light Curves and Detection Probabilities} \label{sec: Simulated Light Curves and Detection Probabilities}
\subsection{Simulated Light Curves}
Our approach is modeled on that of G15, with updated information on the properties of the GBTDS. We used \textit{Kepler} light curves as the basis for the asteroseismic signals. To do so, we selected 100 RC stars and 100 RGB stars from the APOKASC-3 Catalog (\citealt{pinsonneault25}) Gold sample spanning \numax\ values from $\sim 3\ \mu$Hz to $\sim 110\ \mu$Hz to form a representative sample for our simulations. We generated two rank-ordered lists in \numax\ and uniformly sampled the distributions to choose our targets, restricting the list to targets with a full set (18 quarters) of Kepler data. We downloaded \textit{Kepler} light curves for each star through the \texttt{lightkurve} (\citealt{Lightkurve18}) package. We then (1) split each \textit{Kepler} light curve into three 450-day sections, so each would follow the full duration of the GBTDS; (2) adjusted the amplitude of oscillations to account for the change between \textit{Kepler's} bandpass and Roman's F146 wide-filter \citep{Lund19, Sreenivas25}; and (3) injected realistic photometric noise from two different noise models, adjusting for the two assumed cadences (7.5-minute and 15-minute). 

\subsubsection{Amplitude Adjustment}
\label{sec:Amplitude Adjustment}
To make the amplitude adjustment we used the tool \texttt{Gadfly} (\citealt{Gadfly}), which can generate synthetic power spectra by scaling the solar power spectrum given input stellar parameters. In particular, we utilized the \texttt{amplitude\_with\_wavelength} function, which determines the amplitude ratio by integrating a black body spectrum at a given $T_{\rm eff}$ over a specified filter and the SOHO VIRGO PMO6 filter. We obtained an amplitude ratio $A_{\rm F146/Kp}$ by dividing $A_{\rm F146/PM06}$ by $A_{\rm Kp/PM06}$ which are given by the \texttt{amplitude\_with\_wavelength} function over a range of temperatures. The amplitude ratio of the F146 filter over the \textit{Kepler} Kp-band as a function of temperature is described by the following equation:
\begin{equation}
   A_{\rm F146/Kp} = 0.493 + 0.058(T_{5000}) + 0.018(T_{5000})^2,
\label{eq:amp-ratio}
\end{equation}
where $T_{5000} \equiv T_{\rm eff}/5000\ K$. This relationship ranges between $A_{\rm F146}/A_{\rm Kp} \approx 0.540$ and $A_{\rm F146}/A_{\rm Kp} \approx 0.575$ over \teff\ ranging from $3500\ K$ to $5500\ K$.

\subsubsection{Photometric Noise}
\label{sec:Photometric Noise}
\begin{figure}[hbt!]
    \centering
    \includegraphics[width=0.75\linewidth]{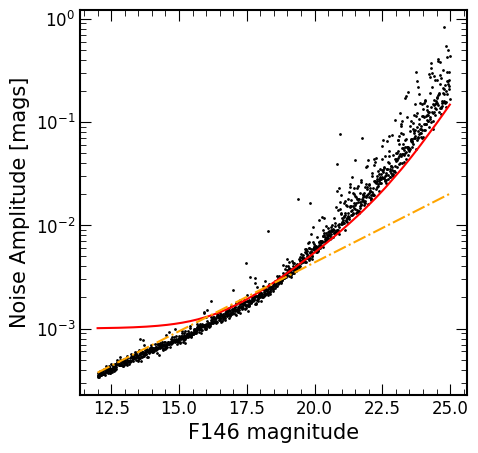}
    \caption{Noise models of the simulations. The red curve shows the Penny noise model, which has a 1 mmag noise floor. The black points show the Wilson noise model, which is based on simulations of saturated star photometry with Roman. The dot-dash orange line shows the noise model described by equation 18 of G15. }
    \label{fig:Noise_models}
\end{figure}

The photon noise floor for the selected
\textit{Kepler} observations is significantly lower than that of the noise floor projected for Roman, so the noise in the \kepler\ light curves can be neglected. We used two models to inject noise at the expected level for Roman into the Kepler light curves. Poisson noise dominates for fainter stars in both, but the first model from \cite{Penny19} (hereafter referred to as the Penny model) assumed a noise floor of $\sim$1mmag for saturated stars. The second model (\citealt{Wislon23}, hereafter referred to as the Wilson model) assumed we can recover more information from saturated stars. The Wilson model was computed by simulating a series of small image cutouts and extracting the uncertainty from each epoch of PSF photometry. The instrument model used to create these simulations utilizes ramp-fitting, but ignores several detector effects that are likely to degrade the quality of observations for stars with brightnesses of F146 $<$ 15--16, such as non-linearities and charge leakage. As a result, the Wilson model is akin to assuming that such effects can be precisely calibrated, which would lead to the photometric noise per pixel being capped at just under the Poisson limit at full well depth. We compare the noise models in Figure \ref{fig:Noise_models}. They diverge brighter than magnitude 16, which is important for our simulations because there is a large population of bulge giants in the 12--15 F146 magnitude range. 

We injected noise into our simulated light curves using both the Wilson and Penny models through the following modified version of equation 19 of G15:
\begin{equation}
F_{\mathrm{F146,}i} = (F_{\mathrm{Kp,}i}-\overline{F_{\rm Kp}})A_{\rm F146/Kp} + N\left(0,\frac{\sigma}{\sqrt{2}}\right),
\label{eq:noise}
\end{equation}
where $F_{\mathrm{Kp,}i}$ is the $i$th observed \textit{Kepler} flux measurement, $\overline{F_{\rm Kp}}$ is the mean of the \textit{Kepler} flux measurements, $A_{\rm F146/Kp}$ is the amplitude ratio defined in equation \ref{eq:amp-ratio}, and $N(x,y)$ is a Gaussian random variable with mean $x$ and variance $y^2$. We injected noise through the $\sigma$ term of equation \ref{eq:noise}, where $\sigma$ is the noise amplitude of a given noise model pictured in Figure \ref{fig:Noise_models}. We included the $\sqrt{2}$ reduction to photometric noise in equation 19 of G15 because the GBTDS will have a nominal cadence of 15 minutes -- half of \textit{Kepler's}. This reduction reflects the shorter integration time per exposure, which reduces the variance of random noise. We further reduced the photometric noise by an additional factor of $\sqrt{2}$ to simulate a two times faster sampling strategy (7.5-minute cadence). 

\begin{figure*}[ht!]
    \includegraphics[width = 1.0\linewidth]{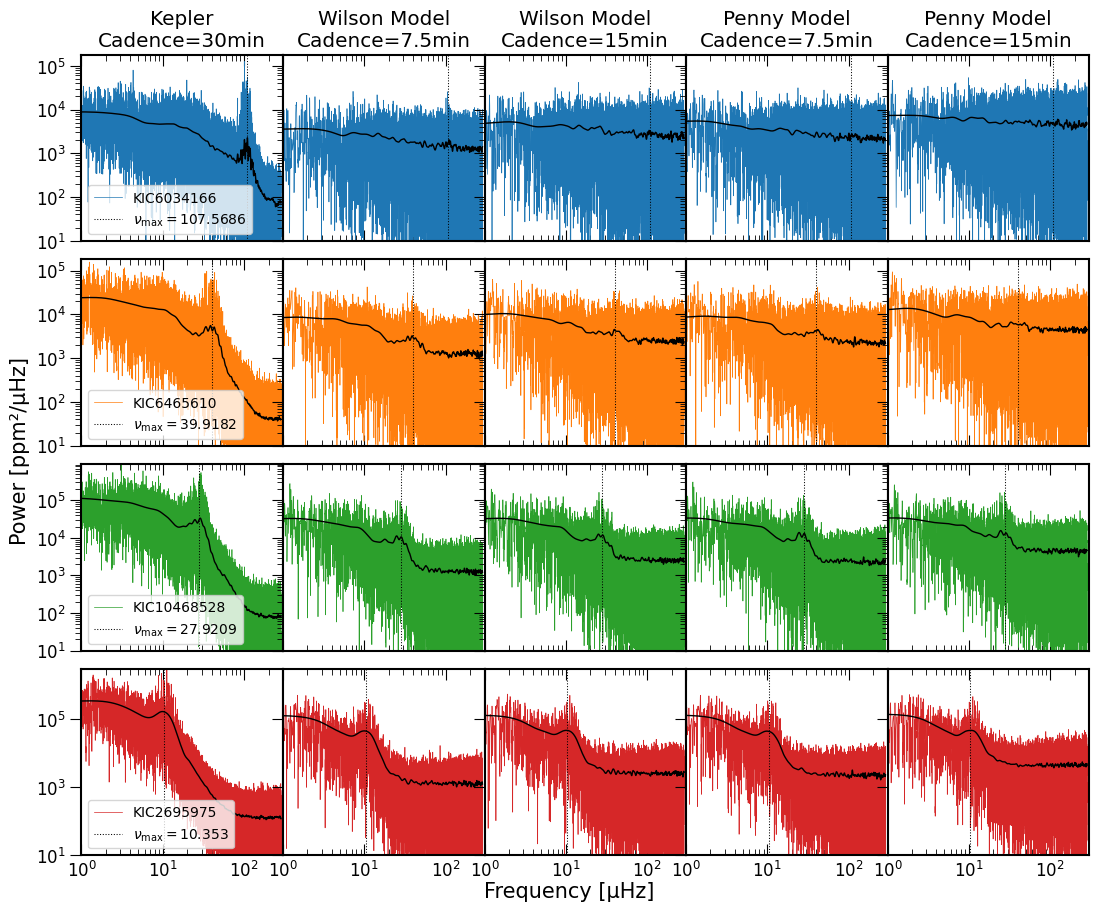}
    \caption{Power spectra of \textit{Kepler} observations and GBTDS simulations as follows (from left to right): \textit{Kepler} 30 minute cadence, Wilson model 7.5 minute cadence, Wilson model 15 minute cadence, Penny model 7.5 minute cadence, and Penny model 15 minute cadence. Each \textit{Kepler} star used for the simulations is labeled with their respective Kepler Input Catalog \citep{Brown2011, 10.17909/t9059r} number (KIC) in the legends of the left-most column. From top to bottom these stars represent the following cases: low-luminosity RGB (KIC 6034166), low-luminosity RC (KIC 6465610), high-luminosity RC (KIC 10468528, and high-luminosity RGB (KIC 2695975). The simulated power spectra were all generated at magnitude 15 in the F146 filter. The colored power spectra are smoothed using a Gaussian filter of width 0.001 $\mu$Hz and the black line shows the power spectra smoothed with a width of 1 $\mu$Hz. The dashed line shows the measured \numax\ for each star in APOKASC3 (\citealt{pinsonneault25}).}
    \label{fig:Simulated_PS}
\end{figure*}

In Figure \ref{fig:Simulated_PS} we plot a representative set of Fourier power spectra of the simulated light curves. The low-luminosity RGB spectrum (top row) is challenging to detect across all noise models, while the more luminous RGB and RC spectra (bottom two rows) are clearly seen in all cases. The lower-luminosity RC spectrum (second row) is located closer to the noise floor. The RC is the main target population, so we can draw two immediate conclusions from this exercise: at least some RC stars should be detectable, and the yields will be sensitive to the noise properties. Fortunately, more luminous RC stars (\numax $\sim25-30\ \mu$Hz) are consistently detectable by eye. It is also seen that a faster cadence can make the oscillation signals more clear. Precise detection probabilities and yields are discussed in more quantitative detail in the following sections.

\subsection{Calculating Signal-to-Noise Ratio and Detection Probability} \label{sec:Detection Probabilities}

Asteroseismology relies on the measurement of \numax\ and \dnu\ to determine stellar parameters as discussed in Section \ref{sec:intro}. Stellar mass can be determined with just \numax\ or just \dnu\ if there is an independent radius measurement. If both are available, an independent radius is not required. With a noise floor close to the oscillation signals, it is important to quantify our ability to measure \numax\ and \dnu, because it is not immediately apparent we will detect oscillations in most cases. In this section we outline how we determine detection probabilities of \numax\ using two methods and how we determine detections of \dnu\ in our simulations. 

Our first method (hereafter referred to as the Chaplin method) computes the signal-to-noise based on the height of the oscillation power excess above the background \citep{Chaplin11}.

\begin{figure*}[ht!]
    \includegraphics[width = 1.0\linewidth]{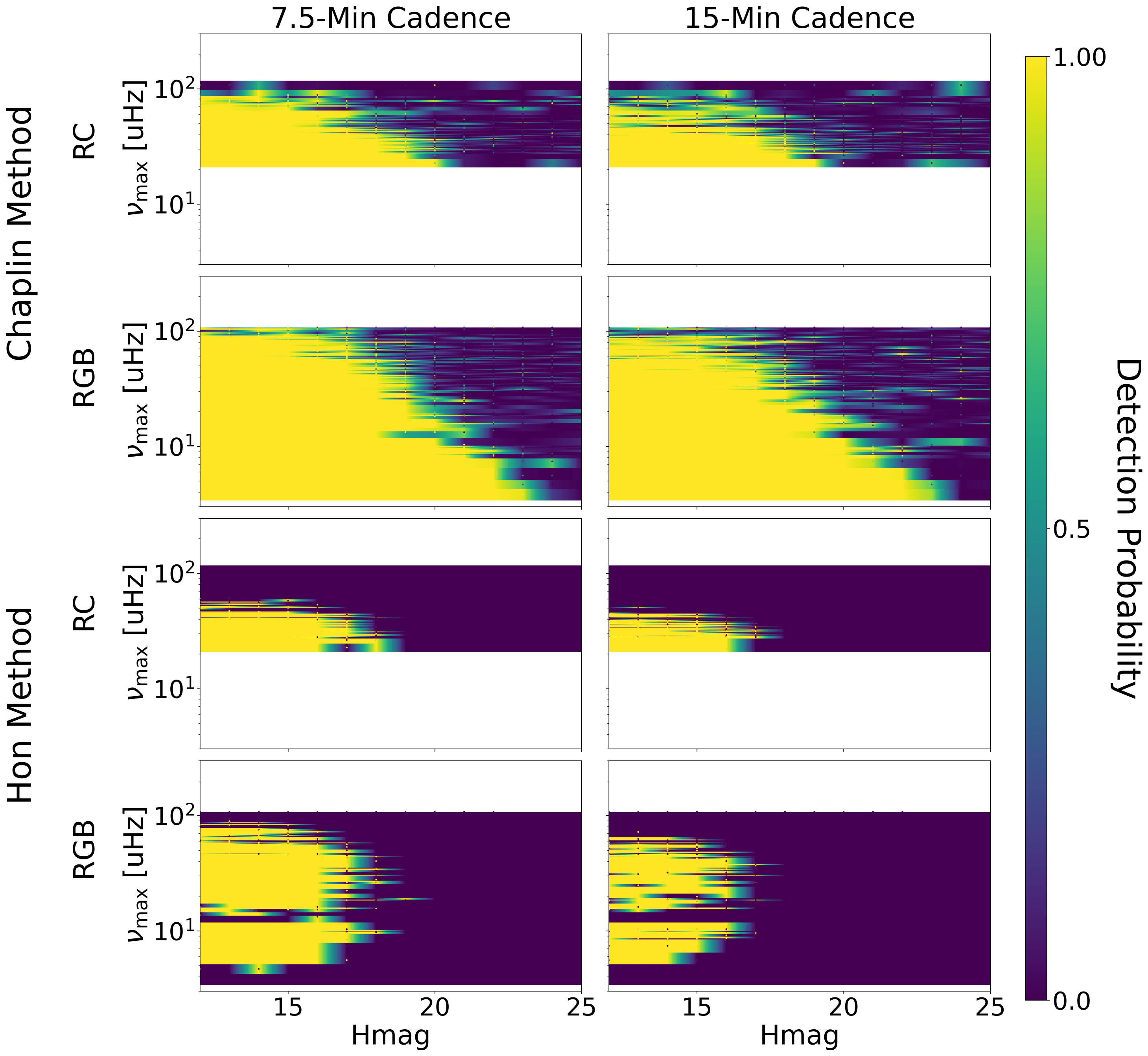}
    \caption{Probability of detection of oscillations as a function of $\nu_{\rm max}$ and H magnitude for different methods and cadences. The top two rows show results using the Chaplin method; the bottom two rows show results from the Hon method. In each pair, the left panel corresponds to a 7.5-minute cadence and the right to a 15-minute cadence. RC (red clump) stars and RGB (red giant branch) stars are shown separately. Brighter regions indicate a higher probability of detection. $\nu_{\rm max}$ values are adopted from APOKASC-3. Individual points represent simulated stars, not Galaxia-generated stars discussed elsewhere. The Hon method exhibits limited detection beyond H $\approx$ 20, reflected in the different detection limits between methods.}
    \label{fig:numax_interpolators}
\end{figure*}

Signal-to-Noise ratios (SNRs) were calculated using the following steps. First we smoothed a given power spectrum using a Gaussian filter of width equal to $\Delta\nu$. Then we inserted a $\pm4\Delta\nu$ gap into the smoothed curve around $\nu_{\rm max}$ and fitted the gap with a straight line in log-space to remove the oscillation signal from the smoothed spectrum. Since the smoothed spectrum then contains only the granulation power and white noise, we calculate the SNR using the following equation
\begin{equation}
\mathrm{SNR} = \frac{1}{N} \sum_{i=1}^{N} \frac{P_i-n_i}{n_i}
\label{eq:snr}
\end{equation}
where $N$ is the number of frequency bins in the power spectrum, $P$ is the raw power spectrum, and $n$ is the smoothed oscillation-free spectrum. We require that the observed SNR is greater than a SNR threshold, $\mathrm{SNR_{thresh}}$, defined using a fractional false-alarm probability of $p=0.01$ which corresponds to the equation
\begin{equation}
    P(\mathrm{SNR'} \ge \mathrm{SNR_{thresh}}, N) = p,
\end{equation}
where $\mathrm{SNR'}$ is an arbitrary SNR and $N$ is the number of frequency bins within $\pm 3\Delta\nu$ around $\nu_{\rm max}$. Then the probability that an observed SNR, $\mathrm{SNR_{tot}}$, is greater than $\mathrm{SNR_{thresh}}$ is given by 
\begin{equation}
   P_{\rm final} = \int_{y}^{\infty} \frac{\exp(-y')}{\Gamma(N)}y'^{(N-1)}dy'
\end{equation}
where $y$ is defined as 
\begin{equation}
    y = (1 + \mathrm{SNR_{thresh}}) / (1+\mathrm{SNR_{tot}}),
\end{equation}
$\Gamma$ is the gamma function, and $\mathrm{SNR_{tot}}$ is determined using Equation \ref{eq:snr}. This gives us a probability of detecting solar-like oscillations. In Figure \ref{fig:numax_interpolators} we visualize the detection probabilities from the Chaplin method using the \numax\--magnitude diagram introduced by \citet{Stello17}.

We also calculated detection probabilities using a pipeline (hereafter referred to as the Hon pipeline) in which oscillations are detected from images of power spectra plotted in log-log space using convolutional neural network classifiers as described in \cite{Hon18}. The classifiers used are similar to that from \cite{Hon19}, in which 4-year \textit{Kepler} power spectra were used as a training set. 
Compared to the Chaplin method, which is a strictly statistical criterion, the Hon pipeline reproduces the detection criteria of the trained eye.
The classifiers directly identify whether the power excess can be detected from observed power spectra. Detection probabilities using the Hon pipeline are visualized in Figure \ref{fig:numax_interpolators}.

\begin{figure*}[ht!]
    \includegraphics[width = 1.0\linewidth]{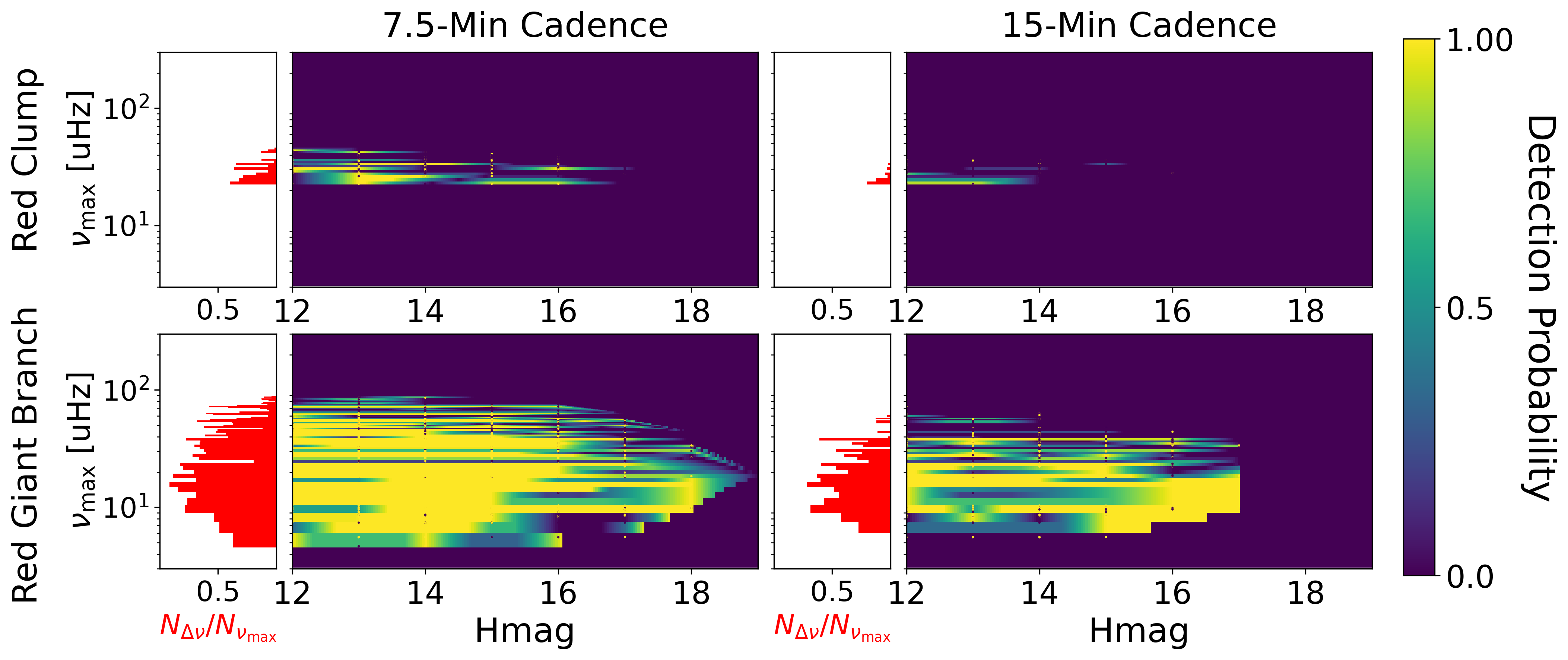}
    \caption{Probability of \dnu\ detection of the nominal case, plotted on \numax\ and H mag, using the SYD pipeline. \numax\ values are adopted from APOKASC-3. The brighter the region, the more likely \dnu\ is detected given  \numax\ and H mag, and the darker the region, the less likely \dnu\ is detected. Note that the visible points represent individual simulated stars, not the stars generated by Galaxia as discussed in later sections. The 1D histogram shows the expected recovery fractions assuming a uniform distribution in magnitude.}
    \label{fig:dnu_Interpolator_Plot}
\end{figure*}

Because the method by \cite{Hon19} only gives the probability of detecting \numax\, we also ran the SYD pipeline \citep{Huber09} to obtain measurements of \dnu, which we vetted using the automated method by \cite{Reyes22}. Detections of \dnu\ are visualized in Figure \ref{fig:dnu_Interpolator_Plot}.

\subsection{Detection Probability Results}
\label{sec:Detection Probability Results}
In Figure \ref{fig:numax_interpolators} we can see the differences in detection probabilities between the Chaplin method and the Hon pipeline, as well as how cadence impacts those probabilities.

The Chaplin method is always more optimistic than the Hon pipeline, but they both agree that \numax\ should be detected in bright RGB and RC stars. This is expected, since the amplitudes drop in less luminous giants, making yields there more sensitive to noise properties. RC stars also have lower amplitude modes compared to RGB stars, explaining the difference in \numax\ detection limits. The Hon pipeline shows that signals may be harder to recover in faint stars than what the formal Chaplin SNR calculation suggests. The Hon method also shows that luminous giants may be harder to recover than the Chaplin method predicts. A decrease in the detection probabilities near \numax $\sim$ 10$\mu$uHz is a feature previously seen in this machine learning Hon detection method (\citealt{Hon21}), which does not perform as well for luminous stars. Were this gap not present, the yields would be larger by $\approx 1\%$, which is comparable to variations in yield estimates due to choice in interpolation methods. Although more noticeable in the Chaplin detection probabilities, both show a diagonal, \numax\ and magnitude-dependent cutoff in detections that reflects the combined effects of (1) decreasing amplitude of oscillation with increasing \numax\ and (2) increasing noise with increasing magnitude. In both methods, implementing a faster sampling speed increases detection probabilities at faint magnitudes for both the RC and RGB, allowing for more detections. Because the Hon method is more conservative overall, we adopt it as our model for detecting \numax. 

\begin{figure}
    \centering
    \includegraphics[width=1.0\linewidth]{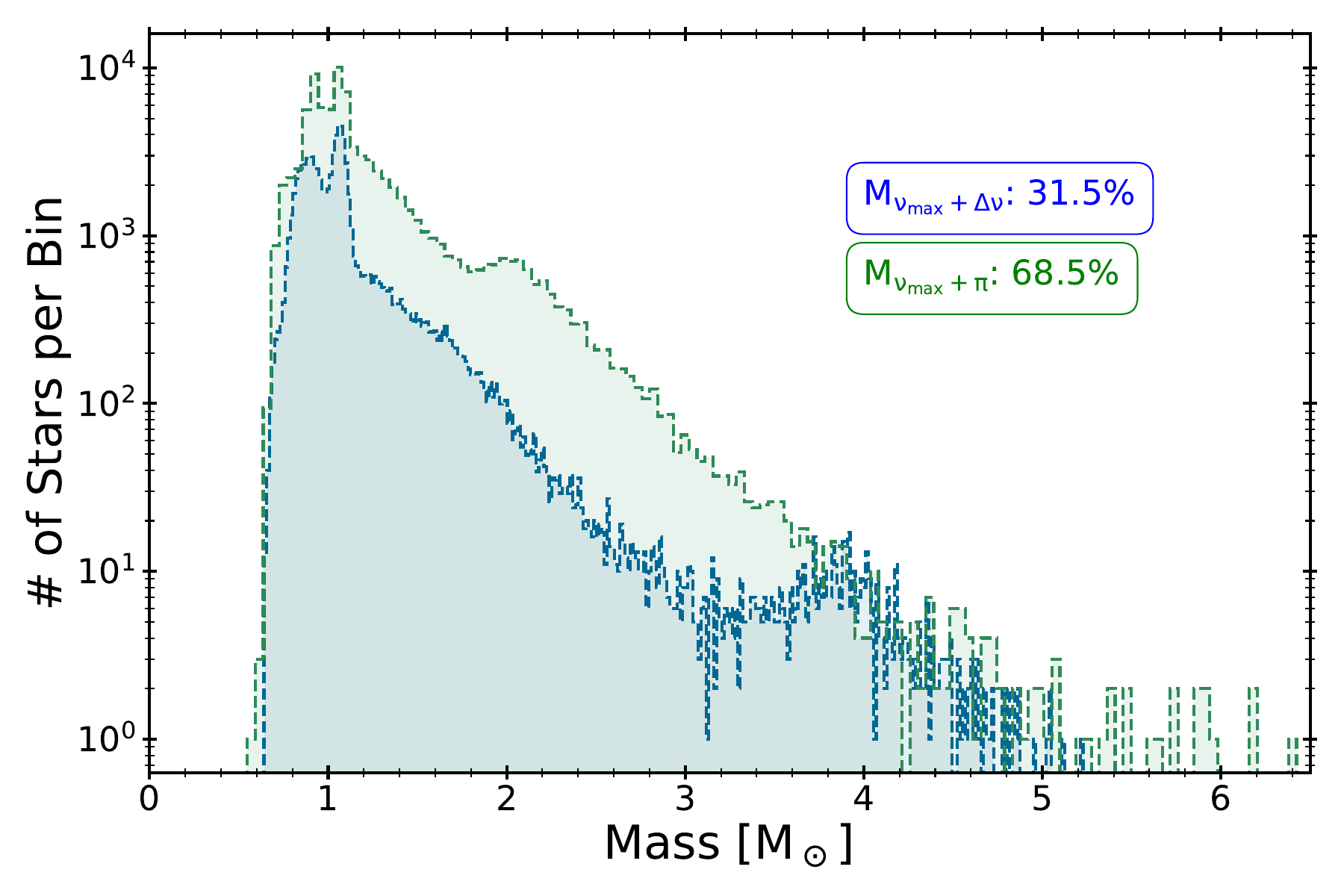}
    \caption{The dark blue histogram shows the distribution of stellar masses for stars which will have detctions in both $\Delta\nu$ and $\nu_{\max}$. When combined with an estimate of $T_{\mathrm{eff}}$, these stars will have inferred stellar masses and radii. The green histrogram shows the distribution of stellar masses for those stars that will have detections in $\nu_{\max}$ only. For these stars, an external estimate of both $T_{\mathrm{eff}}$ and R is needed to estimate the mass.
}
    \label{fig:Mact}
\end{figure}

In Figure \ref{fig:dnu_Interpolator_Plot} we show the probability of detecting \dnu\ in the stars, given a detection of \numax. The sharp drop in probability in the lower right corner of the plots in the bottom row of Figure \ref{fig:dnu_Interpolator_Plot} occurs because there were no stars with a detected \numax\ in those regions, so we input a \dnu\ detection probability of zero. This is also why there are no \dnu\ detections for magnitudes greater than 17 in RGB stars at 15-minute cadence. We find that the probability of detecting \dnu\ is largely independent of magnitude for both the RC and RGB samples, except for a drop near the faint edge of \numax\ detections. We do, however, see a clear \dnu\ detection dependence on \numax. This is because high \numax\ RGB stars exhibit mixed modes making \dnu\ harder to measure (\citealt{Stello13}) and they have smaller oscillation amplitudes, which compounds this issue. Mixed modes are also present in RC stars, which can explain the relatively small \dnu\ detection fractions we see in that population. Mixed mode coupling strengths are several times larger in RC stars than in the RGB sample that have \numax\ detections in our sample; this results in more pronounced mixed-mode appearances in the frequency spectra for RC stars (\citealt{Mosser17}). Coupling strengths are also large at the base of the RGB, but these stars are inaccessible to Roman \numax\ detection. RC stars also generally have overall lower oscillation amplitudes (for the same \numax; \citealt{Yu18}), leading to a more difficult interpretation of the power spectra, which then lowers \dnu\ detections. As seen in the histograms of Figure \ref{fig:dnu_Interpolator_Plot}, a higher cadence leads to higher \dnu\ detection probabilities in regions where \dnu\ was already detected, as well as new detections at high \numax\ where \dnu\ was not detected with lower cadence. This faster cadence affects the signal-to-noise ratio of our objects, increasing the number of detections. The maximum \dnu\ detection fractions, ranging from 21\% for RC stars and 90\% for RGB stars with a 15-minute cadence, are comparable, and even surpass 1-2 sectors of TESS data \citep{Stello22}. This is expected given the longer time series of our Roman GBTDS simulations, leading to higher frequency resolution and sampling of more oscillation cycles. For select figures in this paper (e.g., Figures~\ref{fig:Mact} and~\ref{fig:FeH_Temp_Plot}), a 1\% uncertainty was added to the results purely for visual clarity. This smoothing does not affect the underlying data or alter any scientific findings.

\begin{figure*}[t!]
    \begin{tabular}{p{0.5\linewidth}p{0.5\linewidth}}
    \includegraphics[width = 0.92\linewidth]{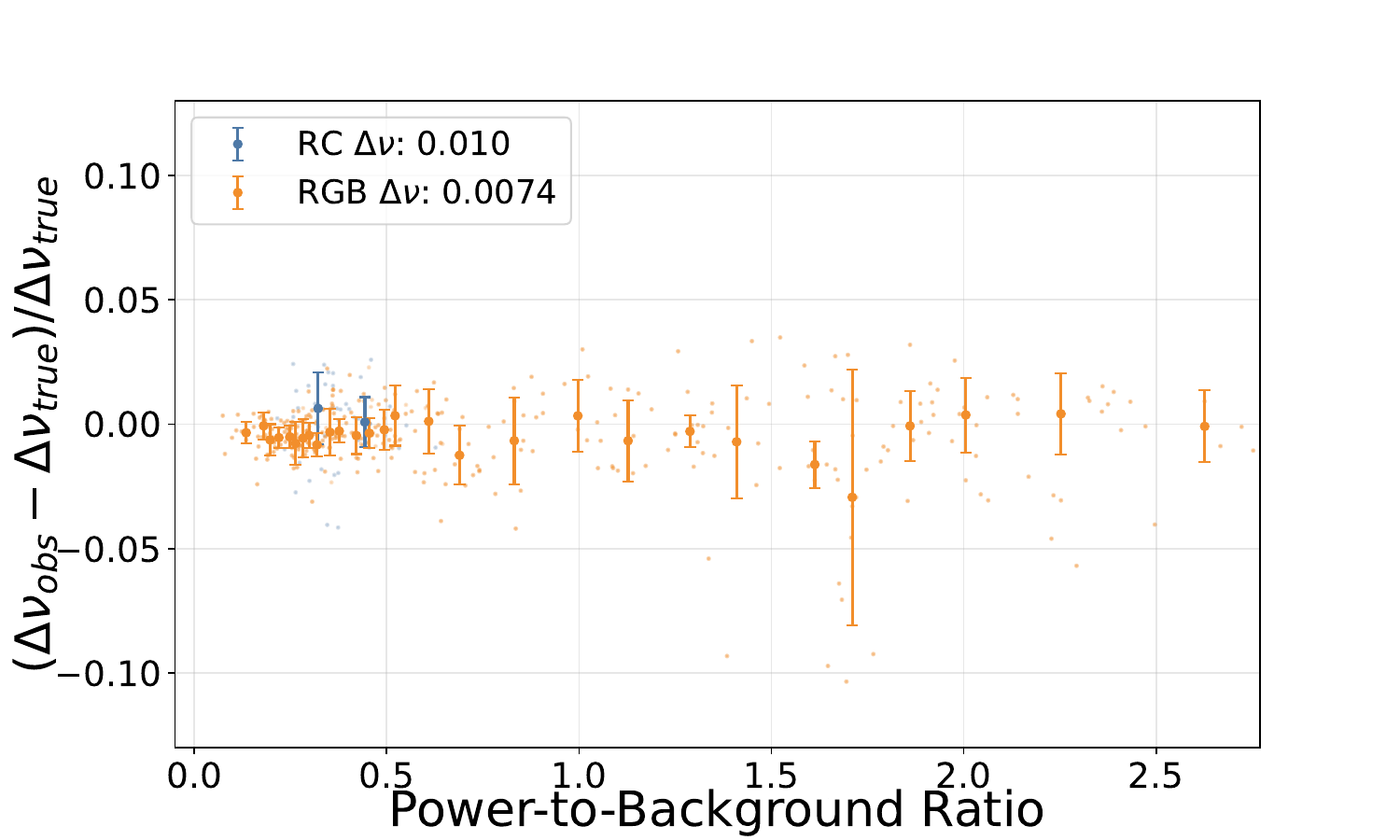}    & \includegraphics[width = 0.92\linewidth]{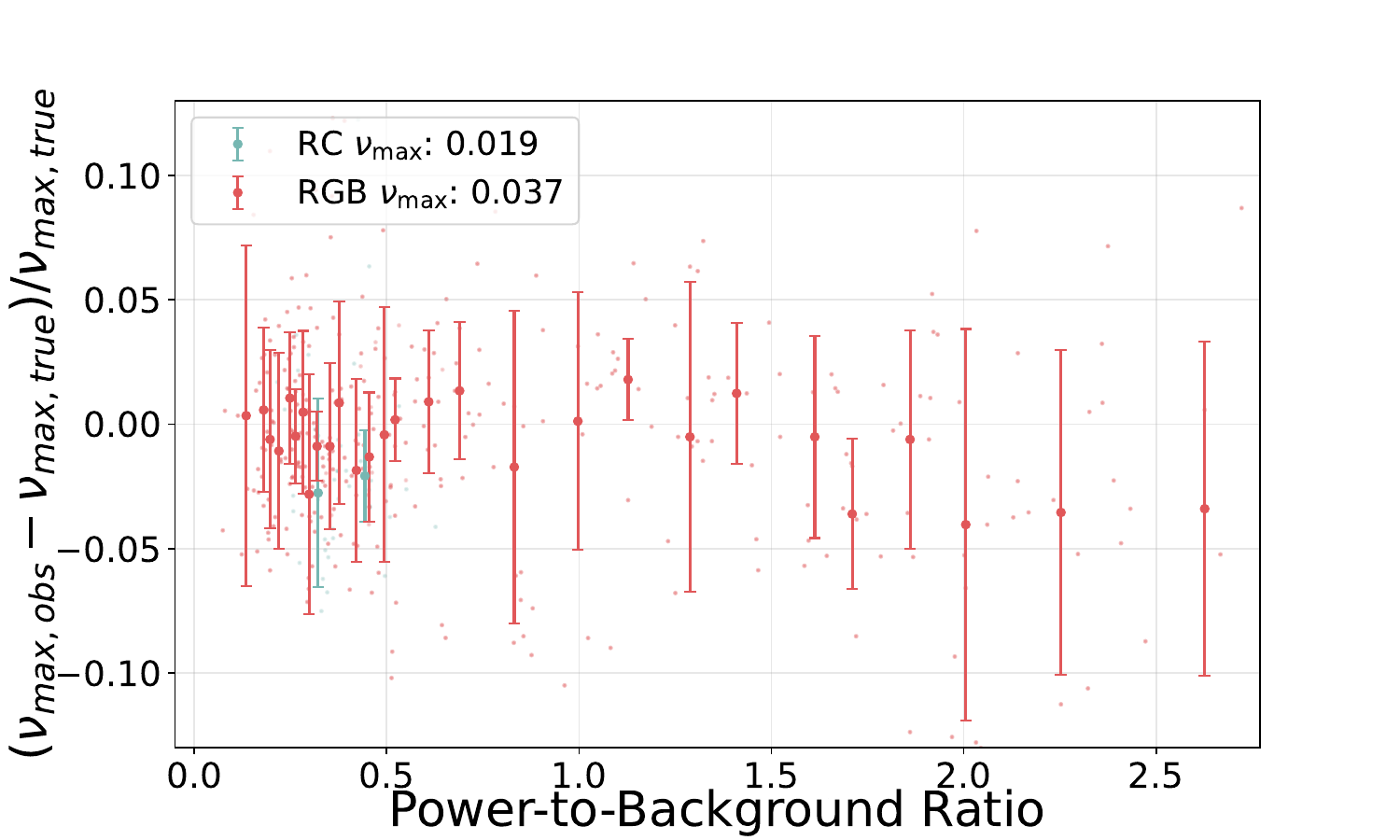}
    \end{tabular}
    \centering
    \caption{\textit{Left}: Fractional deviation of observed \dnu\ relative to the `true'  APOKASC-3 value, as a function of power-to-background ratio, for RC and RGB stars. Error bars are the uncertainties taken from bins, measured by using the median absolute deviation. The uncertainty value used, shown in the legend for RC and RGB, is taken from the bin closest to 0.5 for both cases. \textit{Right}:  Same as the left panel, but instead with observed \numax\ relative to the `true' APOKASC-3 value.}
    \label{fig:uncertainties}
\end{figure*}

 Figure \ref{fig:uncertainties} shows the distribution of the fractional deviation in recovered \numax\ and \dnu\ as a function of power-to-background ratio, which essentially is another means of a signal-to-noise ratio but using calculations based on the Chaplin method. Many of our low \numax\ objects in Figure~\ref{fig:uncertainties} lie in higher power-to-background ratio regions, and can be seen heavily skewing our RGB uncertainties in those bins due to the difficulty of estimating \numax\ at lower frequencies. 
 Note that the majority of the RC detections are found in power-to-background ratios that are less than 0.5, and so do not occupy the same dynamic range of power-to-background ratio as RGB stars in Figure~\ref{fig:uncertainties}.
 
 We use these deviations from `truth' to infer the expected measurement uncertainty using an outlier-insensitive median absolute deviation, 
\begin{equation}
    \text{$\sigma$} = \left[\operatorname{median} \left( \left| X_i - \tilde{X} \right| \right)\right] * 1.4826,
\end{equation}
 where $\tilde{X}$ is the median of the sample, $X_i$ is a sample point, and the constant is a scaling factor. Typical uncertainties for RGB stars at power-to-background ratios of 0.5 are 3.7$\%$ for \numax\  and 0.74$\%$ for \dnu. RC star uncertainties are better for \numax\ at 1.9$\%$, but slightly worsen for \dnu\ at 1.0$\%$. 

We expect approximately $30\%$ of the Roman asteroseismic yield will have both a \dnu\ and a \numax\ detection, shown in Figure \ref{fig:Mact}. In this case, masses can be computed directly via the standard asteroseismic scaling relation (refer to Equation \ref{eq:2}). We note that metallicity and \teff\ are still required to interpret the frequency spacings and to obtain mass, radius, and age. Given the typical uncertainties of \numax\ and \dnu\ we find, the mass uncertainty would be 5.7$\%$ for RC and 8.1$\%$ for RGB, implying age uncertainties of approximately 17$\%$ and 25$\%$ respectively, assuming negligible temperature uncertainties (e.g., $1\%$). With the remaining objects that do not have \dnu\ detections (70$\%$ of our detected sample), we can use the Stefan-Boltzmann law to determine R:
\begin{equation}
    \frac{L}{L_\odot} = \left(\frac{R}{R_\odot}\right)^2\left(\frac{T}{T_\odot}\right)^4,
\end{equation}
where $L$ can be determined with photometric data, an extinction model, and a trigonometric parallax, $\varpi$. In addition to existing parallaxes from Gaia, Roman is expected to deliver precise relative parallaxes for the entire asteroseismic sample, with precisions of 0.3$\mu\mathrm{as}$ \citep{Gould15}. For the typical bulge star, the parallax uncertainty is therefore negligible, and temperature uncertainties dominate. Nevertheless, as yet unknown astrometric systematics may cause the Roman parallaxes to be less precise than the sub-microarcsecond level that was predicted by \citealt{Gould15}.

Combining the radius with a surface gravity from the \numax\ scaling relation then yields a mass, shown in equation \ref{eq:4}. For a temperature uncertainty of 1$\%$, the resulting masses would have a $\sim 4.5\%$ uncertainty due to temperature alone. With typical \numax\ precisions of 2.9$\%$ for the RC, the mass uncertainty comes to $\sim 5.4\%$. For this reason, we expect it will be more advantageous to calculate asteroseismic ages with this mass scale, which would deliver statistical uncertainties in age of closer to 15\% instead of the above-reported 25\% with asteroseismology alone. Note that this does not include uncertainties due to chemical composition or stellar model choice, the latter of which can be particularly significant (e.g. \citealt{Tayar22}).

\section{Asteroseismic Yields} \label{sec: simulated populations}

Having explored the general trends of detection in \numax-\dnu-magnitude space, we now turn to simulated asteroseismic yields using the above detection methods. Our yield calculations depend on both the stellar populations in the GBTDS fields and the line-of-sight extinction. The latter can vary on small spatial scales and the extinction law in the Galactic bulge may not exactly match that of the solar vicinity, though for our purposes we make the assumption that they are the same. The Roman mission will greatly improve our understanding of extinction in these fields. We begin by describing our baseline population model and the associated yields under different detectability scenarios. We then follow up by testing the robustness of our results against changes in the stellar population model.

\subsection{Stellar Population Model}
\label{sec:Stellar Population Model}
 To model the Roman fields, we  generated synthetic stellar populations using \textit{Galaxia} (\citealt{Sharma11}). \textit{Galaxia} creates a synthetic survey of stars in the Milky Way given a field of view and assumed limiting magnitude. Stars are drawn with phase space density consistent with the Besan\c{c}on Milky Way model for the disk (\citealt{Robin03}), including a bar-shaped bulge (\citealt{Blitz93}). The assumed ages for stellar populations in the thin disc vary with metallicity from -0.57 to 0.13. The thick disc is assumed to have an age of 11 Gyr and metallicity of -0.78 $\pm$ 0.3. The bulge is assumed to have an age of 10 Gyr and metallicity of 0.0 $\pm$ 0.4. There is also a halo component, which populates lower metallicites.
 
Individual stars are populated according to Padova isochrones (\citealt{Bertelli94}; \citealt{Marigo08}), with initial mass functions that vary according to the Galactic component (thick disc, bulge, etc.). We refer the reader to \citet{Sharma11} for further details.

 As inputs, we specify the limiting magnitude of the desired survey, the fields of view, as well as a star-by-star model of the extinction.  We then convolve this simulated population with our detection probabilities to infer yields. 
 
 We adopt \textit{Galaxia} because it allows for modification of the input population, such as age-metallicity relations for bulge stars, and it has previously been used to model asteroseismic yields \citep{Sharma16}.\footnote[1]{\url{https://asd.gsfc.nasa.gov/roman/comm_forum/forum_17/Core_Community_Survey_Reports-rev03-compressed.pdf}}

\begin{figure}[hbt!]
    \includegraphics[width = 0.94\linewidth]{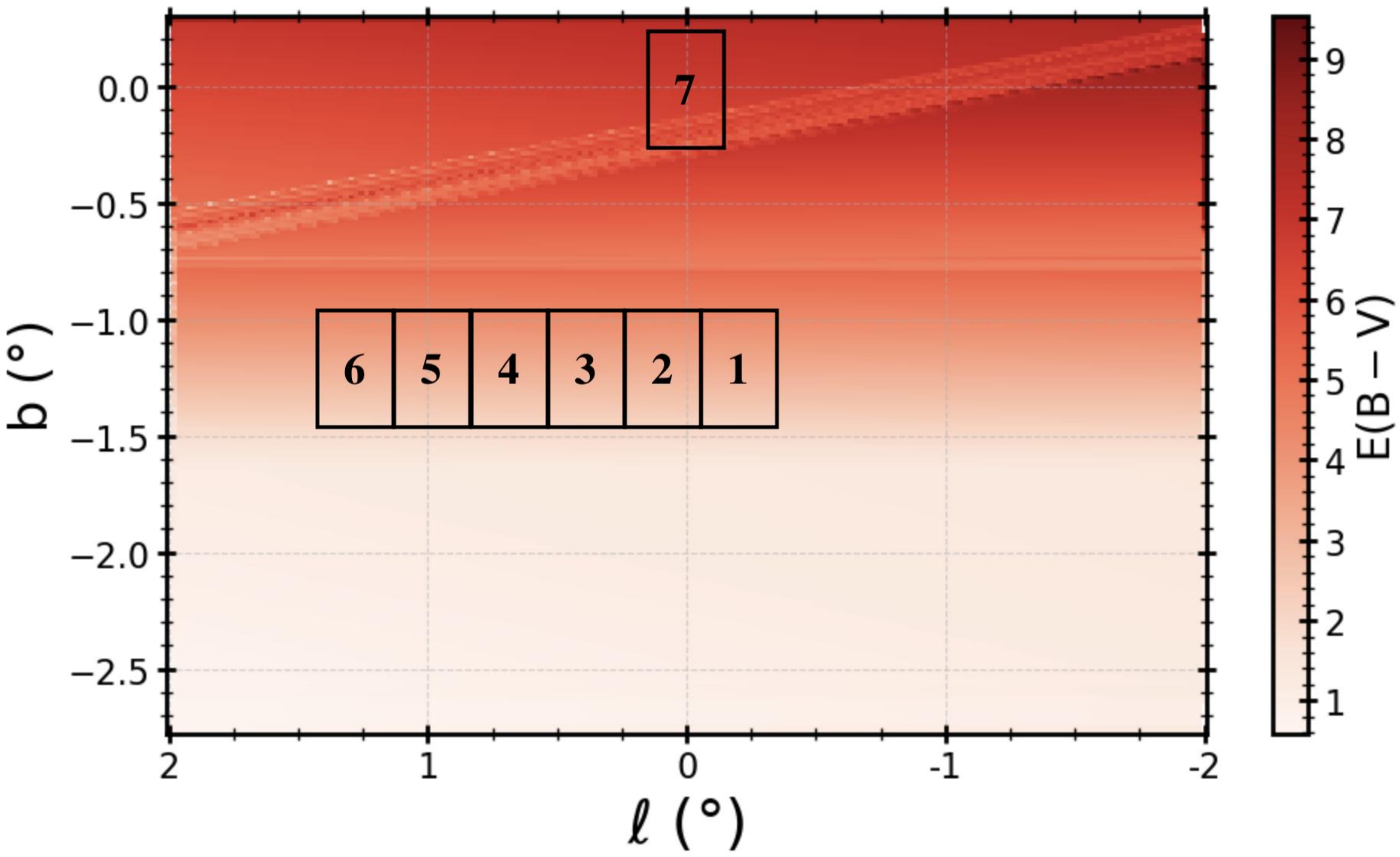}
    \caption{Simulated fields of view for a 15-minute GBTDS strategy, reproduced from the 
    \textit{Roman Galactic Bulge Time-Domain Survey Definition Committee Report}\textsuperscript{1}. 
    Each black polygon represents a Roman pointing. The background represents reddening from  \cite{Marshall06}.}
    \label{fig:FoV}
\end{figure}

 \begin{figure}[hbt!]
    \includegraphics[width = 0.95\linewidth]{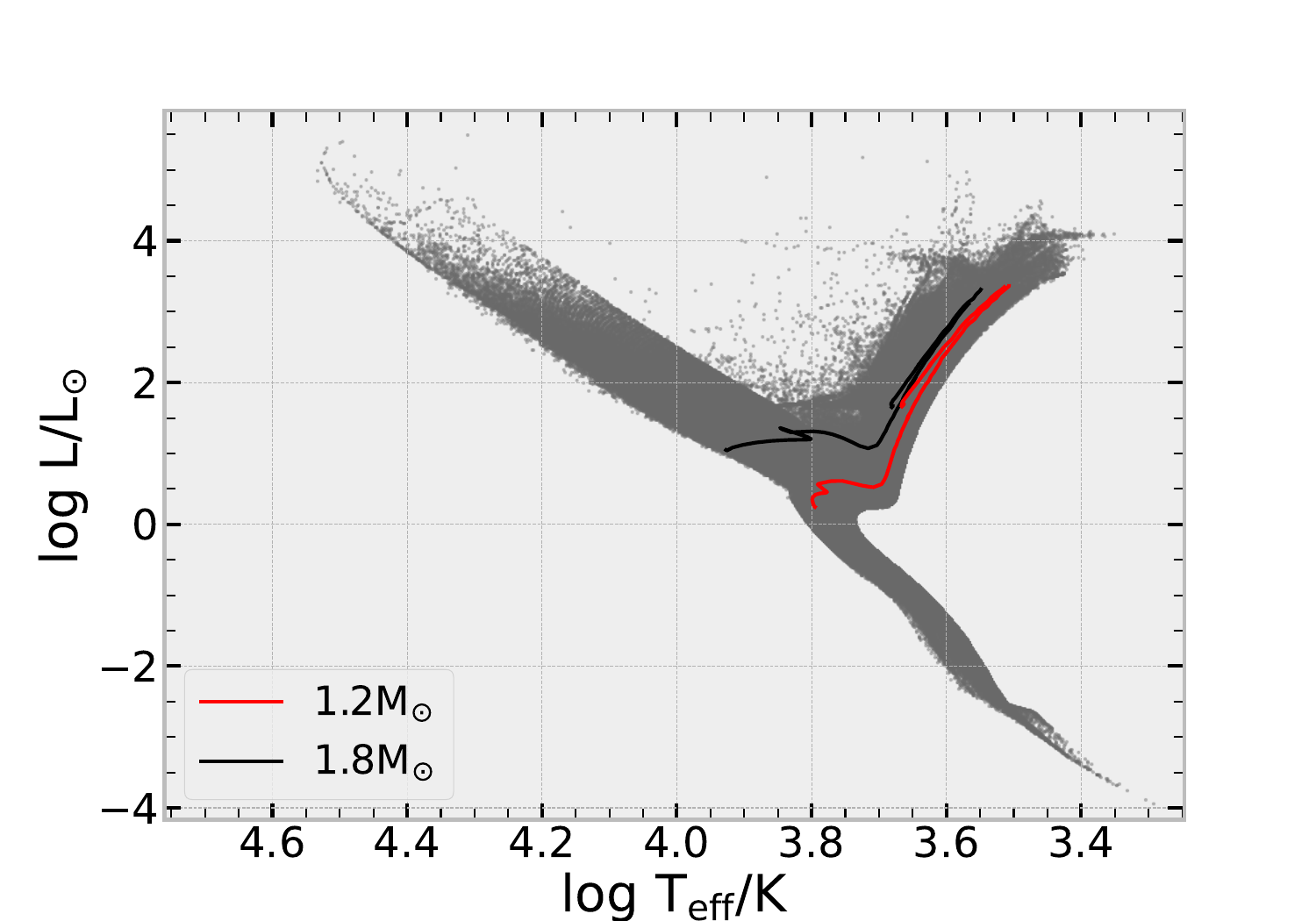}
    \caption{HR diagram of full Galaxia simulation, within nominal field of view set by Roman Core Community suggestions. The red and black lines are evolutionary tracks for different stellar masses, shown in the legend.}
    \label{fig:Galaxia_Plot}
\end{figure}

\subsection{Survey Parameters}
\label{sec:Survey Parameters}
For our \textit{Galaxia} simulations, we adopted fields of view consistent with those proposed in the most recent Roman Core Community report (see Figure \ref{fig:FoV}). The survey is implemented in \textit{Galaxia} using seven circular patches of the sky, each with area ${0.281\ \mathrm{deg}^2}$ (the footprint size of a single Roman pointing, referred in this work as a field of view [FoV]; \citealt{Cromey23}). All the combined footprints, each with area ${0.281\ \mathrm{deg}^2}$, span ${-0.22\degree<l<1.82\degree}$ and ${-1.64\degree<b<-0.85 \degree}$. Note that the simulated fields in \textit{Galaxia} are not quite the correct shape, but this will only have a small effect on our yields. Additionally, there has been recent interest in including a field at the Galactic center (\citealt{Terry23}), so an additional field centered at (0,0) was considered. We used a limiting magnitude of 25, using \textit{2MASS} $H$-band as a proxy for Roman's F146 filter. All magnitudes are AB unless otherwise stated. The entire simulation, in the nominal region and with the listed parameters, produces $\approx$ 18 million objects, seen in Figure \ref{fig:Galaxia_Plot}.

\subsection{Asteroseismic Detection}
\label{sec:Asteroseismic Detection}
 We set asteroseismic detection criteria for objects with ${T_{\rm eff} \le 5250\,\rm K}$, which covers the $T_{\rm eff}$ domain within almost all solar-like oscillators detected in the Kepler and K2 samples. The amplitude depends on \numax\ and the noise properties are sensitive to apparent magnitude, so we account for both in our yields. To do so, we interpolated probabilities, calculated with the Hon method, of the survey, using given $H$-band magnitudes from the survey and calculating \numax\ according to (\citealt{Brown91}, \citealt{Kjeldsen95}):
\begin{equation}
   \nu_{\rm max} =  \nu_{\mathrm{max,\odot}} \left(\frac{g}{g_{\odot}}\right)  
 \left(\frac{T_{\rm eff}}{T_{\mathrm{eff,\odot}}}\right)^{-1/2}.
\end{equation}

We employ \texttt{scipy}'s LinearNDInterpolator (\citealt{SciPy20}), though variations in yields due to interpolator scheme choice are at the percent level. Using the resulting detection probabilities for each star in the \textit{Galaxia} simulation, we then randomly draw a representative `detection' sample.

\subsection{Detection Results}
\label{sec:Detection Results}
We constructed 8 possible detection samples, varying the cadence, detection method, and noise model. In Table \ref{tab:detections} we present the total number of asteroseismic detections, and those of them that belong to the bulge population, for each of the 8 cases.

We select the fourth row (in bold) as our nominal scenario since it is the most conservative estimate. The table can be read as follows: the first column lists the simulated cadence, the second column lists the method used to determine \numax\ detection probabilities, the third column lists the adopted noise model, the fourth column lists the total number of detections, and the fifth column lists the subset of those detections found to belong to the bulge population. 

We adjusted the \textit{Galaxia} models to account for age-metallicity relations recently inferred for the Galactic bulge \citep{Joyce23}. Compared to the nominal \textit{Galaxia} bulge population, this case has a larger proportion of younger ages with higher metallicities. Due to the higher metallicities in the bulge, the resulting detection sample was larger in comparison to the nominal \textit{Galaxia} bulge stellar population, which has more lower-metallicity stars. Lower metallicity RC stars are too hot to support solar-like oscillations, which explains this difference. We discuss this more in \ref{sec:Sample Characteristics}.

For our nominal case, we predict a yield of 290,000 stars, with 185,000 detected belonging to the bulge population. In all cases, the GBTDS is predicted to surpass existing asteroseismic sample sizes. To maintain a conservative approach, we show figures using the nominal case.

\begin{table*}[hbt!]
\centering
\textbf{Specifications of Simulated Detections and Bulge Counts}
\begin{tabular}{|c|c|c|c|c|}
\hline
\textbf{Cadence} & \textbf{Detection Method} & \textbf{Noise Model} & \textbf{Detection Total} & \textbf{Detections in Bulge} \\ \hline
15-Min Cadence & Chaplin & Wilson & 648,000 & 358,000 \\ \hline
15-Min Cadence & Chaplin & Penny & 624,000 & 349,000 \\ \hline
15-Min Cadence & Hon & Wilson & 417,000 & 253,000 \\ \hline
\textbf{15-Min Cadence} & \textbf{Hon} & \textbf{Penny} & \textbf{290,000} & \textbf{185,000} \\ \hline
7.5-Min Cadence & Chaplin & Wilson & 425,000 & 232,000 \\ \hline
7.5-Min Cadence & Chaplin & Penny & 415,000 & 229,000 \\ \hline
7.5-Min Cadence & Hon & Wilson & 342,000 & 195,000 \\ \hline
7.5-Min Cadence & Hon & Penny & 205,000 & 135,000 \\ \hline

\end{tabular}

\caption{A summary of GBTDS cases. `Cadence' refers to how frequently Roman will observe each field (here we simulate a 15- and 7.5-minute cadence). `Detection Method' refers to how we determine \numax\ detection probabilities (either the Chaplin method or Hon pipeline as described in Section \ref{sec:Detection Probabilities}). `Noise Model' refers to the model we adopt for photometric noise (either the Penny model or Wilson model as described in Section \ref{sec:Photometric Noise}). `Detection Total' is the total number of objects detected to have asteroseismology, and `Detections in Bulge' is a subset of those identified to be of the bulge population. The nominal case quoted throughout, and which serves as the basis for figures in the text, is highlighted in bold.} 
\label{tab:detections}
\end{table*}

\begin{figure}[hbt!]
    \includegraphics[width = 1.0\linewidth]{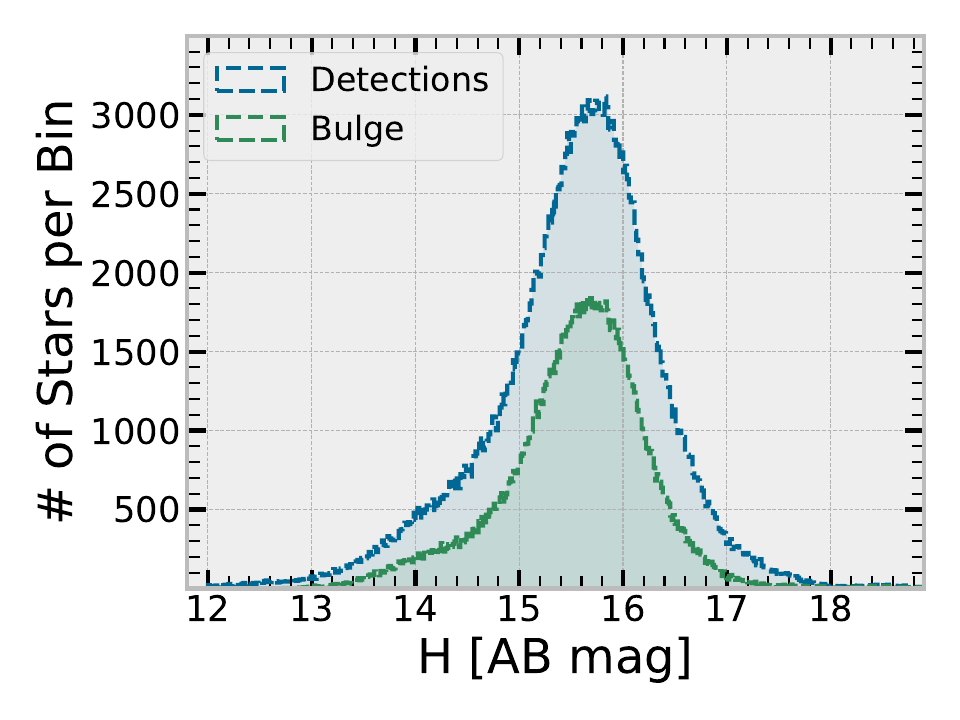}
    \caption{The apparent magnitude distribution of the nominal detection yields nears the saturation limit for Roman ($\approx 15$), and does not extend beyond $\approx18$ due to noise.}
    \label{fig:Hmag_Plot}
\end{figure}

\begin{figure*}[hbt!]
    \includegraphics[width = 1.0\linewidth]{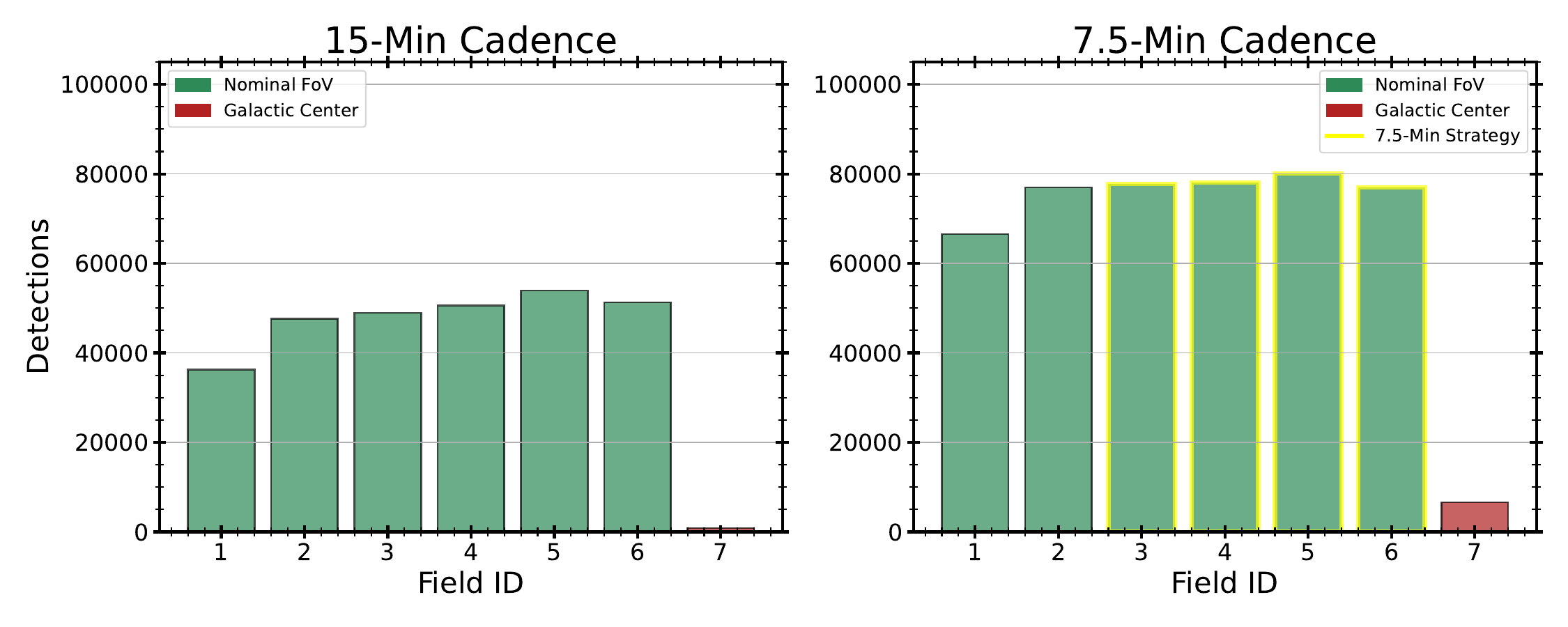}
    \caption{Counts of detections per target field, for both a 15-minute cadence and 7.5-minute cadence of the nominal case. The yellow highlighted bars are the regions that are used in the 7.5-minute cadence survey. Field IDs are as listed in Fig.~\ref{fig:FoV}. Although the 7.5-minute cadence recovers more asteroseismic detections per FoV, it is able to only be used for a limited number of fields, and excludes the Galactic center, as indicated with the yellow highlights. The 15-minute cadence is able to observe in nearly double the number of fields, including the Galactic center.}
    \label{fig:Pop_per_FoV_Histogram}
\end{figure*}

\begin{figure}[hbt!]
    \includegraphics[width = 1.0\linewidth]{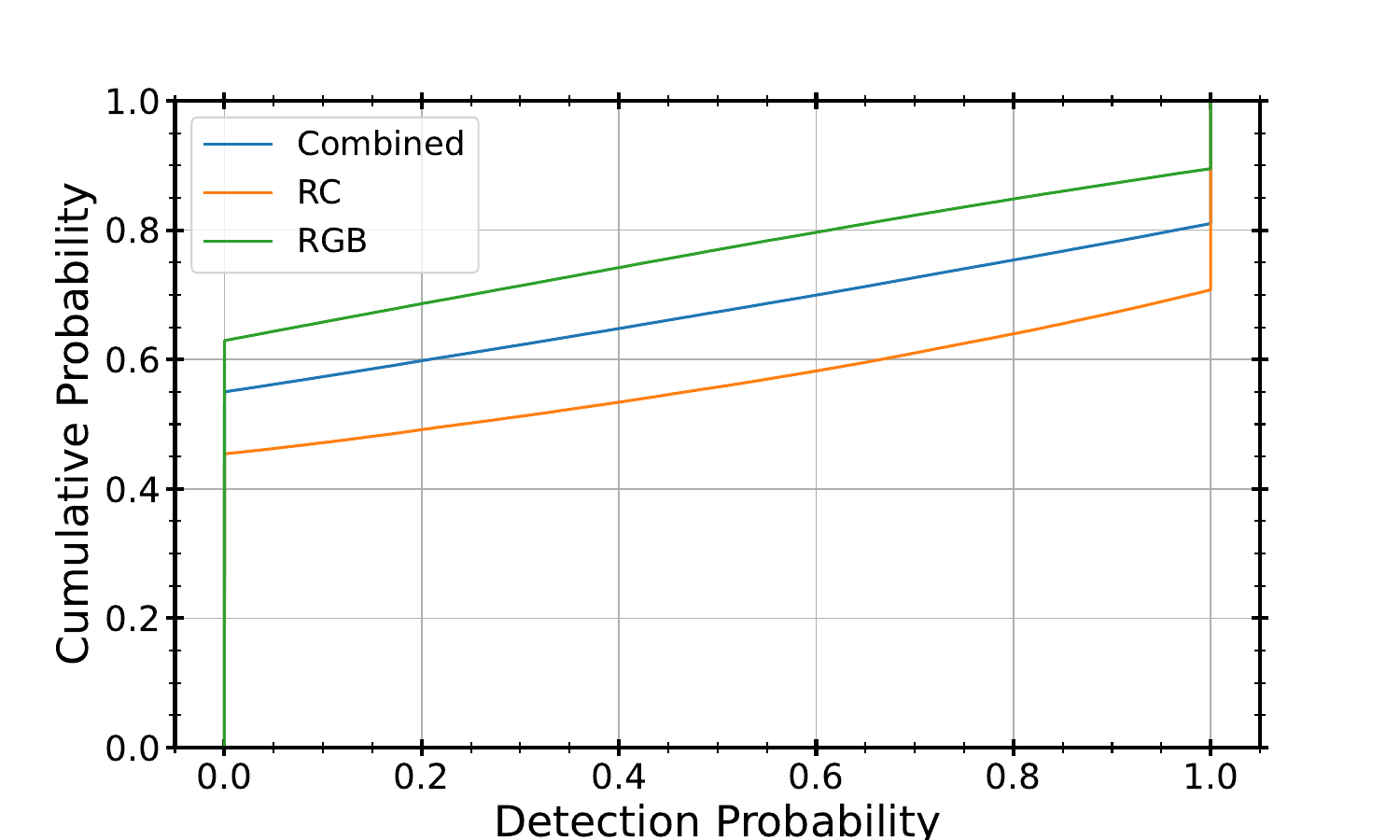}
    \caption{Cumulative distribution plot of the probabilities of detection of the nominal case. This refers to the likelihood (`Cumulative Probability') of an object having a probability (`Detection Probability') to have a measurable asteroseismic signal. The `Combined' function is of the `RC' and `RGB' cumulative plots.}
    \label{fig:CDF_Plot}
\end{figure}

\begin{figure}[hbt!]
    \includegraphics[width = 1.0\linewidth]{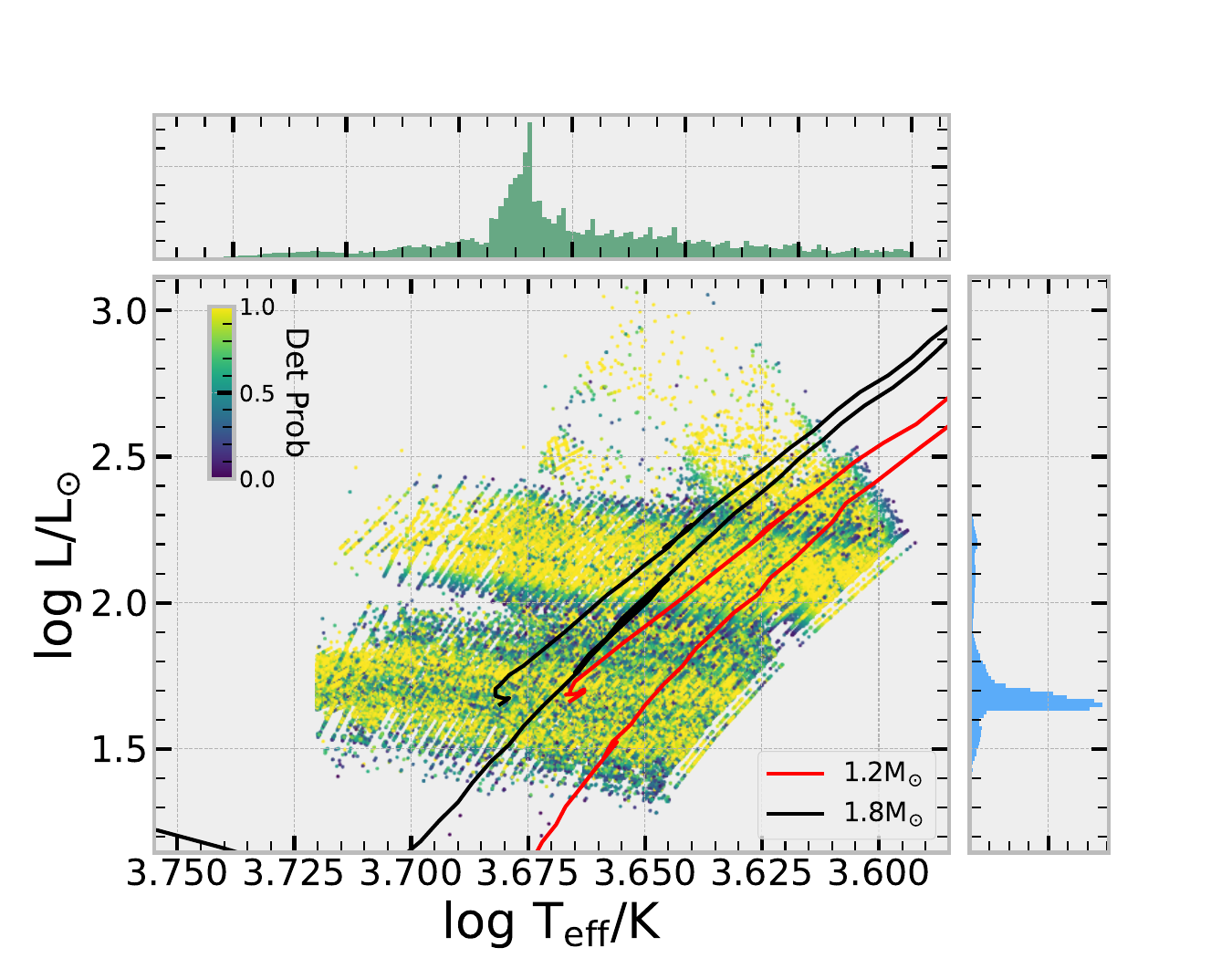}
    \caption{HR diagram of the nominal case, color-coded by the interpolated detection probability. Collapsed histograms show the marginalized temperature and luminosity distributions. Example solar-metallicity Padova evolutionary tracks are provided for reference. The banded structure of the probability detection visible in the figure is primarily due to the variation in detection probability as a function of \numax\ seen in Fig. \ref{fig:numax_interpolators}.}
    \label{fig:HR_Diagram}
\end{figure}

\begin{figure}[hbt!]
    \includegraphics[width = 1.0\linewidth]{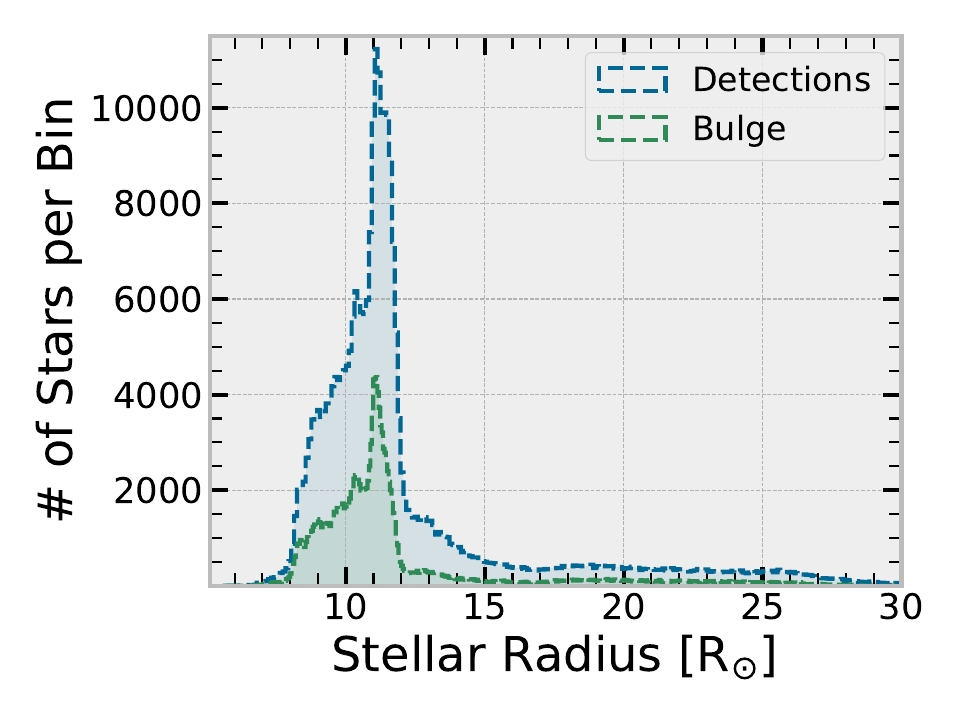}
    \caption{Histogram of both the nominal detections and detections in the bulge as a function of radius, showing that stars at the red clump and below comprise the majority of the expected asteroseismic yields from Roman.}
    \label{fig:Radius_Plot}
\end{figure}

\begin{figure*}[ht!]
    \includegraphics[width = 1.0\linewidth]{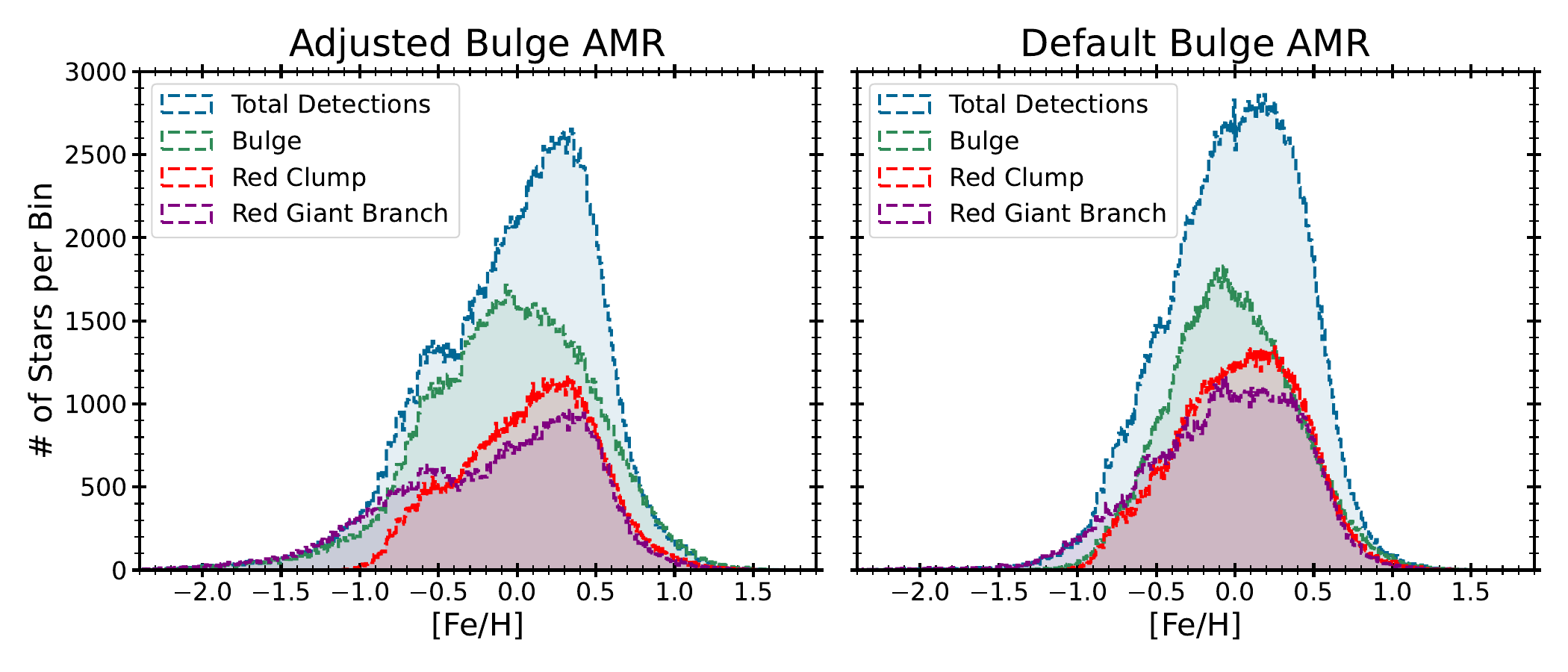}
    \caption{Histogram of the asteroseismic yields with and without modifications to the default \texttt{Galaxia} age-metallicity relation for the bulge, showing modest differences in the resulting metallicity distributions of the yields}.
    \label{fig:FeH_Plot}
\end{figure*}

\begin{figure*}[hbt!]
    \includegraphics[width = 1.0\linewidth]{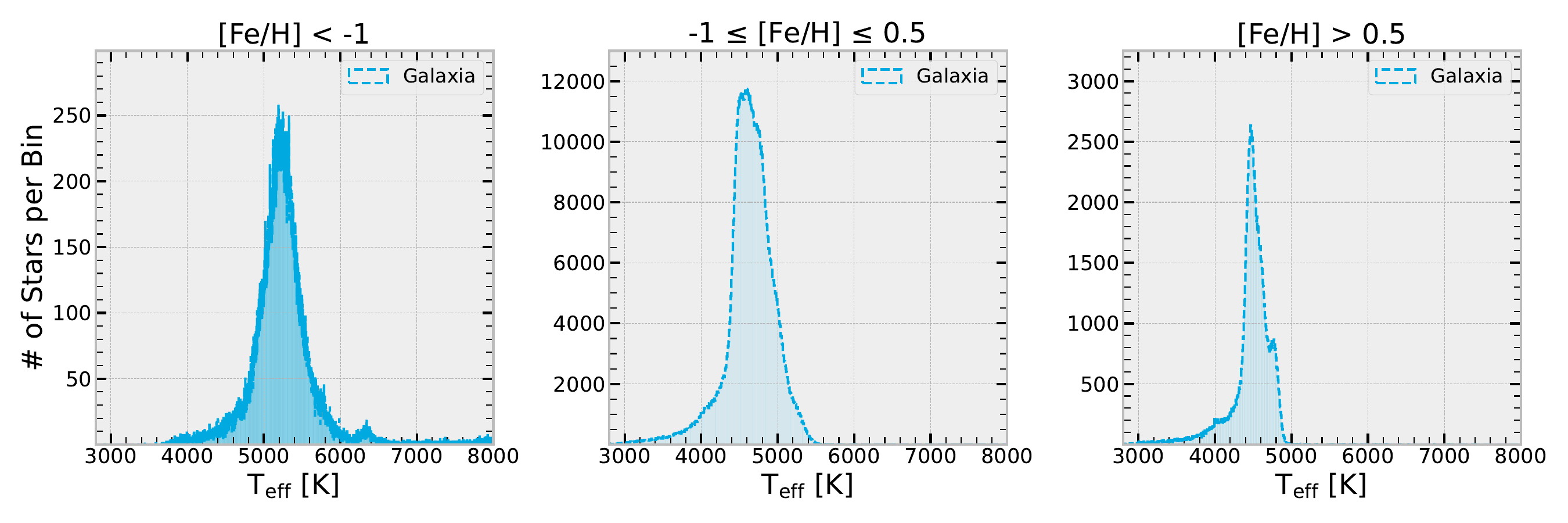}
    \caption{Distributions of simulated \textit{Galaxia} population temperatures for different metallicity bins show why low-metallicity stars are preferentially lost in the asteroseismic yields: their temperatures are largely hotter than our adopted 5250K limit for solar-like oscillations (note the differing y-axis scales). A 1$\%$ uncertainty is applied to the plots.}
    \label{fig:FeH_Temp_Plot}
\end{figure*}

\section{Discussion} \label{sec: Discussion}
Depending on the modeling choices, the detection yields range from approximately 200,000 to nearly 650,000. The following section will discuss these choices in detail and their influence on yield outcomes.

\subsection{Noise Modeling}
We find the selected noise model is important in simulating detections of saturated stars. Roman is likely to perform better than the Penny noise model we present in this paper, but worse than the Wilson model. We see this in Table~\ref{tab:detections}, where the Hon detection method is sensitive to the adopted noise model at the 10--30\% level. The Chaplin method is relatively insensitive to the noise model because even low signal-to-noise cases are deemed detectable.

At the time of writing, the expected performance of saturated stars in Roman is not yet settled, but it seems possible that precise photometry may be possible up to 14th magnitude. As shown in Figure \ref{fig:Hmag_Plot}, the majority of the asteroseismic sample falls between 15th and 16th magnitude, where stars begin to saturate, making this an important consideration.

\subsection{Roman Parameters}
Our yields depend on the specific fields chosen and the observing strategy; neither has been finalized at the time of writing. We have therefore considered different scenarios for evaluating yields. For the 7.5-minute cadence, we assumed that each field would be observed twice in a given cycle, which corresponds to fewer fields but more exposure time per field. For the nominal 15-minute cadence we have more total detections since more fields can be observed, while the 7.5-minute cadence has more detections per field, leading to a more representative sample, as shown in Figure \ref{fig:Pop_per_FoV_Histogram}. From the figure, we see an evident increase of recoverability per FoV from 15-min cadence to 7.5-min cadence. However, due to constraints for the viewing time with Roman, the 7.5-min cadence is limited to only 4 fields (highlighted in yellow), meaning our total count of detections, and in turn our total subset count in the bulge population, is shown to be larger with the 15-min cadence as we are able to observe all fields of view. As shown in Figure \ref{fig:Pop_per_FoV_Histogram}, each field contains a comparable number of simulated objects, with the exception of the Galactic center. Additional simulations across other regions suggest that, provided extinction remains relatively uniform, the stellar density per field remains consistent. Implementing a faster cadence would also improve the quality of  asteroseismic parameters \dnu\ as shown in Figure \ref{fig:dnu_Interpolator_Plot}.

\subsection{Extinction}
When calculating extinction values from dust maps, we found the difference between \cite{Schlegel98} and \cite{Marshall06} only impacts the yields minimally. As mentioned before, Marshall extinction values were used to stay consistent with simulations by \cite{Penny19}. The marginal differences are primarily due to the placement of the fields of view, which happen to be in a region with low extinction. The largest discrepancies in yields are found at the Galactic center, where the most dust is found and therefore the fewest detections. 

\begin{table}[hbt!]
    \begin{center}
        \textbf{Detections Varying Extinction and AMR}
    \end{center}
    \centering
    \begin{tabular}{cc|cc}
        \toprule
        \multicolumn{2}{c|}{\textbf{Schlegel}} & \multicolumn{2}{c}{\textbf{Marshall}} \\
        \cmidrule(lr){1-2} \cmidrule(lr){3-4}
        \textbf{Adjusted} & \textbf{Default} & \textbf{Adjusted} & \textbf{Default} \\
        \midrule
        283,000 & 305,000 & 290,000 & 312,000 \\
        \bottomrule
    \end{tabular}
    \caption{Asteroseismic yields for nominal scenario with \cite{Schlegel98} and \cite{Marshall06}, both with (`Adjusted') and without (`Default') modifications to the bulge age-metallicity relation (AMR). See text for details.}
    \label{tab:AMR}
\end{table}

\subsection{Sample Characteristics} \label{sec:Sample Characteristics}
The nominal asteroseismic detection yields contain a significant number of RC stars. This can be seen with the higher detection probabilities in comparison to the total and RGB in Figure \ref{fig:CDF_Plot} and the resulting prominent peak in the luminosity distribution at $\mathrm{log}\frac{L}{L_{\odot}}\approx 1.7$ in Figure \ref{fig:HR_Diagram}. Conversely, we see that RGB stars have a lower probability of detection, shown in Figure \ref{fig:CDF_Plot}. The RC population is also evident in the radius distribution of the sample, corresponding to a peak of $\sim 11 R_{\odot}$ in Figure \ref{fig:Radius_Plot}.  Nevertheless, RGB stars comprise $40\%$ of the sample. 

With regard to $H$-band distribution, we see a prominent peak just below 16th magnitude (Figure \ref{fig:Hmag_Plot}), which reflects the magnitude of a typical clump star at the center of the bulge. With a faster cadence, we would see a significant number of fainter objects being detected, extending Figure \ref{fig:Hmag_Plot}. If the 7.5-minute cadence strategy included the Galactic center, many of these objects would populate that region, because of the significant extinction at the Galactic center. The 15-minute cadence is less sensitive to regions with higher extinction, making it more difficult to recover fainter objects, especially at the Galactic center. 

All of our simulation cases assume an age-metallicity relation (AMR) consistent with recent work from \cite{Joyce23}. To account for variations in age and metallicity, we simulated three bulge populations, one for each metallicity bin in the \cite{Joyce23} age-metallicity relation, combining them in proportion to the number of objects in each bin. Note that this adjustment applies only to the bulge population. For reference, a case without AMR was also run, using the \texttt{Galaxia} default parameters for the bulge: age of 10 Gyr $\pm$ 0.0 and metallicity of 0.0 [Fe/H] $\pm$ 0.4 for the entire bulge population. The difference in the resulting asteroseismic yields when not assuming \textit{Galaxia}'s default ages and metallicities of the bulge is shown in Figure~\ref{fig:FeH_Plot}, where we see that there are more metal-poor stars in the updated AMR. Nevertheless, the number of metal-poor stars detected remains low because at low metallicity, core-helium burning stars transition to the blue horizontal branch instead of the RC, becoming too hot to exhibit solar-like oscillations. This transition in temperature as a function of metallicity is demonstrated in Figure \ref{fig:FeH_Temp_Plot}.

\section{Conclusion} \label{sec: conclusion}
In this paper, we have simulated realistic time-series data of red giant stars exhibiting solar-like oscillations for the Roman Galactic Bulge Time-Domain Survey (GBTDS). We used Kepler light curves of 100 red clump and 100 RGB stars from APOKASC-3 as a basis for our simulations. We applied modifications to the light curves assuming the GBTDS's nominal 15-minute observing cadence, as well as an additional 7.5-minute cadence, to demonstrate the impact of cadence on yields. We also injected white noise following a conservative noise floor of $\sim1\rm{mmag}$ for saturated stars and a more optimistic noise model with improved saturated star photometry. We derived detection probabilities of the global asteroseismic frequencies, \numax\ and \dnu, by using traditional SNR calculations from \cite{Chaplin11}, a neural-network based asteroseismic detection pipeline \citep{Hon18}, and the SYD pipeline \citep{Huber09}. We then applied these detection probabilities to a simulated survey based on GBTDS strategies using \textit{Galaxia} to model a sample of stellar objects where \numax\ was detected. We further tested our results by varying properties of the Galactic bulge and checking the impact on yields under both the \cite{Schlegel98} and \cite{Marshall06} dust maps and use of a modified age-metallicity relationship described in \cite{Joyce23}. We obtained the following results:
\begin{itemize}
    \item We found that implementing a 7.5-minute cadence can increase yields by up to 50$\%$ per observed field. However, the 7.5-min cadence will have fewer fields compared to the 15-minute cadence, leading to an overall drop in detections, albeit at higher quality.
    \item Since the bulk of our \textit{Galaxia} simulated sample is saturated in the Roman detectors, the photometric noise performance has a significant impact on detections. We found a $\sim 44\%$ increase in total detections compared to our nominal (conservative) case when we implemented a more optimistic noise model.
    \item In contrast, varying the locations of the fields of view and extinction choice had minimal impact on the total number of detected objects.
    \item We found that $\sim 1/3$ of our sample will have both \numax\ and \dnu\ detections, though using \numax\ measurements alone in combination with parallax from Roman and/or Gaia will result in more competitive precisions in mass and age.
    \item Adjusting the age-metallicity relation has a limited, but measurable, affect on the asteroseismic yields.
    \item Given our nominal strategy, we found \textbf{~290,000} total detections, with \textbf{~185,000} of those detections in the Galactic bulge. By varying the selected cadence, noise model, and detection algorithm, we found total detections as low as 205,000 and as high as 648,000, corresponding to bulge detections of 135,000 and 358,000, respectively.
\end{itemize}

In future work, we plan on cross-matching our simulated results with data collected by \textit{Gaia} (\citealt{Gaia16}, \citealt{Gaia23}) and \textit{2MASS} (\citealt{MASS06}), which will be useful in constructing an asteroseismic target list in preparation for Roman's launch. The stellar population simulations presented here complement this effort by providing an estimate of the completeness of the Roman asteroseismic yields, which will be crucial for exoplanet-hosting population characterization and studies of the age distributions of bulge stars.

Seeing that it is possible to measure \numax\ for a large number of stars and \dnu\ for a subset of that population, it is necessary to determine our ability to measure $T_{\rm eff}$ and $R$ in the bulge. Roman multiband photometry, grism spectroscopy, and precise astrometry could be useful in this regard (G15; Roman Core Community report).

Our simulated yields prove the transformative potential of asteroseismology with the Roman GBTDS. Stellar ages with competitive uncertainties will be available for the first time for hundreds of thousands of red giants in the bulge, and should enable a number of stellar population and Galactic archaeology applications. 

\section*{Acknowledgments}
This work was supported by NASA award 80NSSC24K0091. This work was partially supported by funding from the Department of Astronomy of the Ohio State University. This paper includes data collected by the Kepler mission and obtained from the MAST data archive at the Space Telescope Science Institute (STScI). We thank Matthew Penny and Jennifer Johnson for their contributions and advisement throughout the duration of the work. We thank Tom Barclay, Joshua Schlieder, and the Roman Team for their insight on the project. We acknowledge support from the Australian Research Council for D.S. (DP190100666) and T.R.B. (FL220100117). B.S.G. was supported by the Thomas Jefferson Chair for Discovery and Space Exploration Endowment.

\software{Galaxia (\citealt{Sharma11}), 
SciPy (\citealt{SciPy20}),
astropy (\citealt{astropy13, astropy18, astropy22}), 
mwdust (\citealt{Bovy16}),
LightKurve (\citealt{Lightkurve18}),
pandas (\citealt{pandas20}),
Matplotlib (\citealt{Hunter07}),
seaborn (\citealt{Waskom21}),
numpy (\citealt{Harris20})
}

\newpage
\bibliographystyle{aasjournal}
\bibliography{bibliography.bib}

\providecommand{\noopsort}[1]{}
\begin{thebibliography}{}
\expandafter\ifx\csname natexlab\endcsname\relax\def\natexlab#1{#1}\fi
\providecommand{\url}[1]{\href{#1}{#1}}
\providecommand{\dodoi}[1]{doi:~\href{http://doi.org/#1}{\nolinkurl{#1}}}
\providecommand{\doeprint}[1]{\href{http://ascl.net/#1}{\nolinkurl{http://ascl.net/#1}}}
\providecommand{\doarXiv}[1]{\href{https://arxiv.org/abs/#1}{\nolinkurl{https://arxiv.org/abs/#1}}}

\bibitem[{Abbott {et~al.}(2017)Abbott, Valluri, Shen, {et~al.}}]{Abbott17}
Abbott, C.~G., Valluri, M., Shen, J., {et~al.} 2017, MNRAS, 470, 1526, \dodoi{10.1093/mnras/stx1262}

\bibitem[{{Ash} {et~al.}(2025){Ash}, {Pinsonneault}, {Vrard}, \& {Zinn}}]{Ash25}
{Ash}, A.~L., {Pinsonneault}, M.~H., {Vrard}, M., \& {Zinn}, J.~C. 2025, \apj, 979, 135, \dodoi{10.3847/1538-4357/ad9b18}

\bibitem[{{Astropy Collaboration} {et~al.}(2013){Astropy Collaboration}, {Robitaille}, {Tollerud}, {Greenfield}, {Droettboom}, {Bray}, {Aldcroft}, {Davis}, {Ginsburg}, {Price-Whelan}, {Kerzendorf}, {Conley}, {Crighton}, {Barbary}, {Muna}, {Ferguson}, {Grollier}, {Parikh}, {Nair}, {Unther}, {Deil}, {Woillez}, {Conseil}, {Kramer}, {Turner}, {Singer}, {Fox}, {Weaver}, {Zabalza}, {Edwards}, {Azalee Bostroem}, {Burke}, {Casey}, {Crawford}, {Dencheva}, {Ely}, {Jenness}, {Labrie}, {Lim}, {Pierfederici}, {Pontzen}, {Ptak}, {Refsdal}, {Servillat}, \& {Streicher}}]{astropy13}
{Astropy Collaboration}, {Robitaille}, T.~P., {Tollerud}, E.~J., {et~al.} 2013, \aap, 558, A33, \dodoi{10.1051/0004-6361/201322068}

\bibitem[{{Astropy Collaboration} {et~al.}(2018){Astropy Collaboration}, {Price-Whelan}, {Sip{\H{o}}cz}, {G{\"u}nther}, {Lim}, {Crawford}, {Conseil}, {Shupe}, {Craig}, {Dencheva}, {Ginsburg}, {Vand erPlas}, {Bradley}, {P{\'e}rez-Su{\'a}rez}, {de Val-Borro}, {Aldcroft}, {Cruz}, {Robitaille}, {Tollerud}, {Ardelean}, {Babej}, {Bach}, {Bachetti}, {Bakanov}, {Bamford}, {Barentsen}, {Barmby}, {Baumbach}, {Berry}, {Biscani}, {Boquien}, {Bostroem}, {Bouma}, {Brammer}, {Bray}, {Breytenbach}, {Buddelmeijer}, {Burke}, {Calderone}, {Cano Rodr{\'\i}guez}, {Cara}, {Cardoso}, {Cheedella}, {Copin}, {Corrales}, {Crichton}, {D'Avella}, {Deil}, {Depagne}, {Dietrich}, {Donath}, {Droettboom}, {Earl}, {Erben}, {Fabbro}, {Ferreira}, {Finethy}, {Fox}, {Garrison}, {Gibbons}, {Goldstein}, {Gommers}, {Greco}, {Greenfield}, {Groener}, {Grollier}, {Hagen}, {Hirst}, {Homeier}, {Horton}, {Hosseinzadeh}, {Hu}, {Hunkeler}, {Ivezi{\'c}}, {Jain}, {Jenness}, {Kanarek}, {Kendrew}, {Kern}, {Kerzendorf}, {Khvalko}, {King}, {Kirkby}, {Kulkarni},
  {Kumar}, {Lee}, {Lenz}, {Littlefair}, {Ma}, {Macleod}, {Mastropietro}, {McCully}, {Montagnac}, {Morris}, {Mueller}, {Mumford}, {Muna}, {Murphy}, {Nelson}, {Nguyen}, {Ninan}, {N{\"o}the}, {Ogaz}, {Oh}, {Parejko}, {Parley}, {Pascual}, {Patil}, {Patil}, {Plunkett}, {Prochaska}, {Rastogi}, {Reddy Janga}, {Sabater}, {Sakurikar}, {Seifert}, {Sherbert}, {Sherwood-Taylor}, {Shih}, {Sick}, {Silbiger}, {Singanamalla}, {Singer}, {Sladen}, {Sooley}, {Sornarajah}, {Streicher}, {Teuben}, {Thomas}, {Tremblay}, {Turner}, {Terr{\'o}n}, {van Kerkwijk}, {de la Vega}, {Watkins}, {Weaver}, {Whitmore}, {Woillez}, {Zabalza}, \& {Astropy Contributors}}]{astropy18}
{Astropy Collaboration}, {Price-Whelan}, A.~M., {Sip{\H{o}}cz}, B.~M., {et~al.} 2018, \aj, 156, 123, \dodoi{10.3847/1538-3881/aabc4f}

\bibitem[{{Astropy Collaboration} {et~al.}(2022){Astropy Collaboration}, {Price-Whelan}, {Lim}, {Earl}, {Starkman}, {Bradley}, {Shupe}, {Patil}, {Corrales}, {Brasseur}, {N{"o}the}, {Donath}, {Tollerud}, {Morris}, {Ginsburg}, {Vaher}, {Weaver}, {Tocknell}, {Jamieson}, {van Kerkwijk}, {Robitaille}, {Merry}, {Bachetti}, {G{"u}nther}, {Aldcroft}, {Alvarado-Montes}, {Archibald}, {B{'o}di}, {Bapat}, {Barentsen}, {Baz{'a}n}, {Biswas}, {Boquien}, {Burke}, {Cara}, {Cara}, {Conroy}, {Conseil}, {Craig}, {Cross}, {Cruz}, {D'Eugenio}, {Dencheva}, {Devillepoix}, {Dietrich}, {Eigenbrot}, {Erben}, {Ferreira}, {Foreman-Mackey}, {Fox}, {Freij}, {Garg}, {Geda}, {Glattly}, {Gondhalekar}, {Gordon}, {Grant}, {Greenfield}, {Groener}, {Guest}, {Gurovich}, {Handberg}, {Hart}, {Hatfield-Dodds}, {Homeier}, {Hosseinzadeh}, {Jenness}, {Jones}, {Joseph}, {Kalmbach}, {Karamehmetoglu}, {Ka{l}uszy{'n}ski}, {Kelley}, {Kern}, {Kerzendorf}, {Koch}, {Kulumani}, {Lee}, {Ly}, {Ma}, {MacBride}, {Maljaars}, {Muna}, {Murphy}, {Norman}, {O'Steen},
  {Oman}, {Pacifici}, {Pascual}, {Pascual-Granado}, {Patil}, {Perren}, {Pickering}, {Rastogi}, {Roulston}, {Ryan}, {Rykoff}, {Sabater}, {Sakurikar}, {Salgado}, {Sanghi}, {Saunders}, {Savchenko}, {Schwardt}, {Seifert-Eckert}, {Shih}, {Jain}, {Shukla}, {Sick}, {Simpson}, {Singanamalla}, {Singer}, {Singhal}, {Sinha}, {Sip{H{o}}cz}, {Spitler}, {Stansby}, {Streicher}, {{{S}}umak}, {Swinbank}, {Taranu}, {Tewary}, {Tremblay}, {Val-Borro}, {Van Kooten}, {Vasovi{'c}}, {Verma}, {de Miranda Cardoso}, {Williams}, {Wilson}, {Winkel}, {Wood-Vasey}, {Xue}, {Yoachim}, {Zhang}, {Zonca}, \& {Astropy Project Contributors}}]{astropy22}
{Astropy Collaboration}, {Price-Whelan}, A.~M., {Lim}, P.~L., {et~al.} 2022, \apj, 935, 167, \dodoi{10.3847/1538-4357/ac7c74}

\bibitem[{{Belkacem} {et~al.}(2011){Belkacem}, {Goupil}, {Dupret}, {Samadi}, {Baudin}, {Noels}, \& {Mosser}}]{Belkacem11}
{Belkacem}, K., {Goupil}, M.~J., {Dupret}, M.~A., {et~al.} 2011, \aap, 530, A142, \dodoi{10.1051/0004-6361/201116490}

\bibitem[{Bensby {et~al.}(2017)Bensby, Feltzing, Gould, {et~al.}}]{Bensby17}
Bensby, T., Feltzing, S., Gould, A., {et~al.} 2017, A\&A, 605, 34, \dodoi{10.1051/0004-6361/201730560}

\bibitem[{Berger {et~al.}(2020)Berger, Huber, Gaidos, {et~al.}}]{Berger20}
Berger, T., Huber, D., Gaidos, E., {et~al.} 2020, The Astronomical Journal, 160, 108B, \dodoi{10.3847/1538-3881/aba18a}

\bibitem[{Bertelli {et~al.}(1994)Bertelli, Bressan, Chiosi, {et~al.}}]{Bertelli94}
Bertelli, G., Bressan, A., Chiosi, C., {et~al.} 1994, Astronomy \& Astrophysics Suppl., 106, 275

\bibitem[{Blitz {et~al.}(1993)Blitz, Binney, Lo, {et~al.}}]{Blitz93}
Blitz, L., Binney, J., Lo, K., {et~al.} 1993, Nature, 361, 417, \dodoi{10.1038/361417a0}

\bibitem[{Borucki {et~al.}(2010)Borucki, Koch, Basri, {et~al.}}]{Borucki10}
Borucki, W., Koch, D., Basri, G., {et~al.} 2010, Science, 327, 977, \dodoi{10.1126/science.1185402}

\bibitem[{Borucki {et~al.}(1997)Borucki, Koch, Dunham, \& Jenkins}]{Borucki97}
Borucki, W., Koch, D., Dunham, E., \& Jenkins, J. 1997, ASP, 119, 153, \dodoi{1997ASPC..119..153B}

\bibitem[{Bovy {et~al.}(2016)Bovy, Rix, Green, {et~al.}}]{Bovy16}
Bovy, J., Rix, H.-W., Green, G.~M., {et~al.} 2016, ApJ, 818, 130, \dodoi{10.3847/0004-637X/818/2/130}

\bibitem[{Brown {et~al.}(1991)Brown, Gilliland, Noyes, \& Ramsey}]{Brown91}
Brown, T., Gilliland, R., Noyes, R., \& Ramsey, L. 1991, ApJ, 368, 599, \dodoi{10.1086/169725}

\bibitem[{{Brown} {et~al.}(2011){Brown}, {Latham}, {Everett}, \& {Esquerdo}}]{Brown2011}
{Brown}, T.~M., {Latham}, D.~W., {Everett}, M.~E., \& {Esquerdo}, G.~A. 2011, \aj, 142, 112, \dodoi{10.1088/0004-6256/142/4/112}

\bibitem[{Chaplin {et~al.}(2011)Chaplin, Kjeldsen, Bedding, {et~al.}}]{Chaplin11}
Chaplin, W., Kjeldsen, H., Bedding, T., {et~al.} 2011, ApJ, 732, 9, \dodoi{10.1088/0004-637X/732/1/54}

\bibitem[{Chaplin \& Miglio(2013)}]{Chaplin13}
Chaplin, W., \& Miglio, A. 2013, ARAA, 51, 353, \dodoi{10.1146/annurev-astro-082812-140938}

\bibitem[{Ciucă {et~al.}(2021)Ciucă, Kawata, Miglio, {et~al.}}]{Ciucă21}
Ciucă, I., Kawata, D., Miglio, A., {et~al.} 2021, MNRAS, 503, 2814, \dodoi{10.1093/mnras/stab639}

\bibitem[{Cromey {et~al.}(2023)Cromey, Handorf, Pedroncelli, {et~al.}}]{Cromey23}
Cromey, B., Handorf, R.~V., Pedroncelli, J., {et~al.} 2023, Proceeding of the SPIE, 12676, 12 pp, \dodoi{10.1117/12.2676488}

\bibitem[{David {et~al.}(2021)David, Contardo, Sandoval, {et~al.}}]{David21}
David, T., Contardo, G., Sandoval, A., {et~al.} 2021, The Astronomical Journal, 161, 265, \dodoi{10.3847/1538-3881/abf439}

\bibitem[{{Gaia Collaboration} {et~al.}(2016){Gaia Collaboration}, Brown, Vallenari, {et~al.}}]{Gaia16}
{Gaia Collaboration}, Brown, A. G.~A., Vallenari, A., {et~al.} 2016, A\&A, 595, 23 pp, \dodoi{10.1051/0004-6361/201629512}

\bibitem[{{Gaia Collaboration} {et~al.}(2023){Gaia Collaboration}, Vallenari, Brown, {et~al.}}]{Gaia23}
{Gaia Collaboration}, Vallenari, A., Brown, A. G.~A., {et~al.} 2023, A\&A, 674, 22 pp, \dodoi{10.1051/0004-6361/202243940}

\bibitem[{Gilliland {et~al.}(2010)Gilliland, Brown, Christensen-Dalsgaard, {et~al.}}]{Gilliland10}
Gilliland, R., Brown, T., Christensen-Dalsgaard, J., {et~al.} 2010, PASP, 122, 131, \dodoi{10.1086/650399}

\bibitem[{Girardi(2016)}]{Girardi16}
Girardi, L. 2016, ARAA, 54, 95, \dodoi{10.1146/annurev-astro-081915-023354}

\bibitem[{Gould {et~al.}(2015)Gould, Huber, Penny, \& Stello}]{Gould15}
Gould, A., Huber, D., Penny, M., \& Stello, D. 2015, JKAS, 48, 93, \dodoi{10.5303/JKAS.2015.48.2.93}

\bibitem[{Harris {et~al.}(2020)Harris, Millman, van~der Walt, Gommers, Virtanen, Cournapeau, Wieser, Taylor, Berg, Smith, Kern, Picus, Hoyer, van Kerkwijk, Brett, Haldane, del R{\'{i}}o, Wiebe, Peterson, G{\'{e}}rard-Marchant, Sheppard, Reddy, Weckesser, Abbasi, Gohlke, \& Oliphant}]{Harris20}
Harris, C.~R., Millman, K.~J., van~der Walt, S.~J., {et~al.} 2020, Nature, 585, 357, \dodoi{10.1038/s41586-020-2649-2}

\bibitem[{Hayden {et~al.}(2015)Hayden, Bovy, Holtzman, {et~al.}}]{Hayden15}
Hayden, M., Bovy, J., Holtzman, J., {et~al.} 2015, ApJ, 808, 18, \dodoi{10.1088/0004-637X/808/2/132}

\bibitem[{{Hekker}(2020)}]{Hekker20}
{Hekker}, S. 2020, Frontiers in Astronomy and Space Sciences, 7, 3, \dodoi{10.3389/fspas.2020.00003}

\bibitem[{Hey {et~al.}(2023)Hey, Huber, Shappee, {et~al.}}]{Hey23}
Hey, D., Huber, D., Shappee, B., {et~al.} 2023, AJ, 166, 12, \dodoi{10.3847/1538-3881/ad01bf}

\bibitem[{Hon {et~al.}(2021)Hon, Huber, Kuszlewicz, {et~al.}}]{Hon21}
Hon, M., Huber, D., Kuszlewicz, J.~S., {et~al.} 2021, ApJ, 919, 18, \dodoi{10.3847/1538-4357/ac14b1}

\bibitem[{Hon {et~al.}(2019)Hon, Stello, García, {et~al.}}]{Hon19}
Hon, M., Stello, D., García, R., {et~al.} 2019, MNRAS, 485, 5616, \dodoi{10.1093/mnras/stz622}

\bibitem[{Hon {et~al.}(2018)Hon, Stello, \& Yu}]{Hon18}
Hon, M., Stello, D., \& Yu, J. 2018, MNRAS, 476, 3233, \dodoi{10.1093/mnras/sty483}

\bibitem[{Huber {et~al.}(2023)Huber, Pinsionneault, Beck, {et~al.}}]{Huber23}
Huber, D., Pinsionneault, M., Beck, P., {et~al.} 2023, eprint arXiv, \dodoi{10.48550/arXiv.2307.03237}

\bibitem[{Huber {et~al.}(2009)Huber, Stello, Bedding, {et~al.}}]{Huber09}
Huber, D., Stello, D., Bedding, T., {et~al.} 2009, CoAst, 160, 74, \dodoi{10.48550/arXiv.0910.2764}

\bibitem[{Hunter(2007)}]{Hunter07}
Hunter, J.~D. 2007, Computing in Science \& Engineering, 9, 90, \dodoi{10.1109/MCSE.2007.55}

\bibitem[{Jackiewicz(2021)}]{Jackiewicz21}
Jackiewicz, J. 2021, Frontiers in Astronomy and Space Sciences, 7, 102, \dodoi{10.3389/fspas.2020.595017}

\bibitem[{Joyce {et~al.}(2023)Joyce, Johnson, Marchetti, {et~al.}}]{Joyce23}
Joyce, M., Johnson, C.~I., Marchetti, T., {et~al.} 2023, ApJ, 946, 31, \dodoi{10.3847/1538-4357/acb692}

\bibitem[{Kallinger {et~al.}(2010)Kallinger, Weiss, Barban, {et~al.}}]{Kallinger10}
Kallinger, T., Weiss, W., Barban, C., {et~al.} 2010, A\&A, 509, A77, \dodoi{10.1051/0004-6361/200811437}

\bibitem[{Kjeldsen \& Bedding(1995)}]{Kjeldsen95}
Kjeldsen, H., \& Bedding, T. 1995, A\&A, 293, 87, \dodoi{10.48550/arXiv.astro-ph/9403015}

\bibitem[{Kurtz(2022)}]{Kurtz22}
Kurtz, D. 2022, ARAA, 60, 31, \dodoi{10.1146/annurev-astro-052920-094232}

\bibitem[{Leung {et~al.}(2023)Leung, Bovy, Mackereth, {et~al.}}]{Leung23}
Leung, H.~W., Bovy, J., Mackereth, J.~T., {et~al.} 2023, MNRAS, 522, 4577, \dodoi{10.1093/mnras/stad1272}

\bibitem[{Li {et~al.}(2023)Li, Bedding, Stello, {et~al.}}]{Li-Yaguang23}
Li, Y., Bedding, T.~R., Stello, D., {et~al.} 2023, MNRAS, 523, 916, \dodoi{10.1093/mnras/stad1445}

\bibitem[{LightkurveCollaboration(2018)}]{Lightkurve18}
LightkurveCollaboration. 2018, Astrophysics Source Code Library, \dodoi{2018ascl.soft12013L}

\bibitem[{{Lund}(2019)}]{Lund19}
{Lund}, M.~N. 2019, \mnras, 489, 1072, \dodoi{10.1093/mnras/stz2010}

\bibitem[{MacKereth {et~al.}(2019)MacKereth, Bovy, Leung, {et~al.}}]{Mackereth19}
MacKereth, J.~T., Bovy, J., Leung, H.~W., {et~al.} 2019, MNRAS, 489, 176, \dodoi{10.1093/mnras/stz1521}

\bibitem[{Marigo {et~al.}(2008)Marigo, Girardi, Bressan, {et~al.}}]{Marigo08}
Marigo, P., Girardi, L., Bressan, A., {et~al.} 2008, Astronomy \& Astrophysics, 482, 883, \dodoi{10.1051/0004-6361:20078467}

\bibitem[{Marshall {et~al.}(2006)Marshall, Robin, Reylé, {et~al.}}]{Marshall06}
Marshall, D.~J., Robin, A.~C., Reylé, C., {et~al.} 2006, A\&A, 453, 635, \dodoi{10.1051/0004-6361:20053842}

\bibitem[{Martig {et~al.}(2016)Martig, Fouesneau, Rix, {et~al.}}]{Martig16}
Martig, M., Fouesneau, M., Rix, H.-W., {et~al.} 2016, MNRAS, 456, 3655, \dodoi{10.1093/mnras/stv2830}

\bibitem[{Miglio {et~al.}(2021)Miglio, Chiappini, Mackereth, {et~al.}}]{Miglio21}
Miglio, A., Chiappini, C., Mackereth, J.~T., {et~al.} 2021, A\&A, 645, 24, \dodoi{10.1051/0004-6361/202038307}

\bibitem[{Miglio {et~al.}(2013)Miglio, Chiappini, Morel, {et~al.}}]{Miglio13}
Miglio, A., Chiappini, C., Morel, T., {et~al.} 2013, MNRAS, 429, 423, \dodoi{10.1093/mnras/sts345}

\bibitem[{Morris \& Huber(in prep)}]{Gadfly}
Morris, B., \& Huber, D. in prep, \dodoi{https://github.com/bmorris3/gadfly?tab=readme-ov-file}

\bibitem[{Mosser {et~al.}(2017)Mosser, Pin\c{c}on, Belkacem, {et~al.}}]{Mosser17}
Mosser, B., Pin\c{c}on, C., Belkacem, K., {et~al.} 2017, A\&A, 600, 10, \dodoi{10.1051/0004-6361/201630053}

\bibitem[{{Nascimbeni} {et~al.}(2022){Nascimbeni}, {Piotto}, {B{\"o}rner}, {Montalto}, {Marrese}, {Cabrera}, {Marinoni}, {Aerts}, {Altavilla}, {Benatti}, {Claudi}, {Deleuil}, {Desidera}, {Fabrizio}, {Gizon}, {Goupil}, {Granata}, {Heras}, {Magrin}, {Malavolta}, {Mas-Hesse}, {Ortolani}, {Pagano}, {Pollacco}, {Prisinzano}, {Ragazzoni}, {Ramsay}, {Rauer}, \& {Udry}}]{2022A&A...658A..31N}
{Nascimbeni}, V., {Piotto}, G., {B{\"o}rner}, A., {et~al.} 2022, \aap, 658, A31, \dodoi{10.1051/0004-6361/202142256}

\bibitem[{Nataf(2015)}]{Nataf15}
Nataf, D.~M. 2015, Astronomical Society of the Pacific Conference Series, 491, 174

\bibitem[{Ness {et~al.}(2016)Ness, Hogg, Rix, {et~al.}}]{Ness16b}
Ness, M., Hogg, D.~W., Rix, H.-W., {et~al.} 2016, ApJ, 823, 19 pp., \dodoi{10.3847/0004-637X/823/2/114}

\bibitem[{Ness \& Lang(2016)}]{Ness16a}
Ness, M., \& Lang, D. 2016, The Astronomical Journal, 152, 4 pp., \dodoi{10.3847/0004-6256/152/1/14}

\bibitem[{pandas~development team(2020)}]{pandas20}
pandas~development team, T. 2020, pandas-dev/pandas: Pandas, latest,  Zenodo, \dodoi{10.5281/zenodo.3509134}

\bibitem[{Penny {et~al.}(2019)Penny, Gaudi, Kerins, {et~al.}}]{Penny19}
Penny, M.~T., Gaudi, B., Kerins, E., {et~al.} 2019, ApJ Supplement Series, 241, 3, \dodoi{10.3847/1538-4365/aafb69}

\bibitem[{Pinsonneault {et~al.}(2025)Pinsonneault, Zinn, Tayar, {et~al.}}]{pinsonneault25}
Pinsonneault, M., Zinn, J., Tayar, J., {et~al.} 2025, ApJ Supplement Series, \dodoi{10.3847/1538-4365/ad9fef}

\bibitem[{Pinsonneault {et~al.}(2018)Pinsonneault, Elsworth, Tayar, {et~al.}}]{Pinsonneault18}
Pinsonneault, M.~H., Elsworth, Y.~P., Tayar, J., {et~al.} 2018, ApJ Supplemental Series, 239, 32 pp., \dodoi{10.3847/1538-4365/aaebfd}

\bibitem[{Reyes {et~al.}(2022)Reyes, Stello, Hon, {et~al.}}]{Reyes22}
Reyes, C., Stello, D., Hon, M., {et~al.} 2022, MNRAS, 511, 5578, \dodoi{10.1093/mnras/stac445}

\bibitem[{Robin {et~al.}(2003)Robin, Reylé, Derrière, {et~al.}}]{Robin03}
Robin, A.~C., Reylé, C., Derrière, S., {et~al.} 2003, A\&A, 409, 523, \dodoi{10.1051/0004-6361:20031117}

\bibitem[{Schlegel {et~al.}(1998)Schlegel, Finkbeiner, \& Davis}]{Schlegel98}
Schlegel, D.~J., Finkbeiner, D.~P., \& Davis, M. 1998, ApJ, 500, 525, \dodoi{10.1086/305772}

\bibitem[{Sharma {et~al.}(2011)Sharma, Bland-Hawthorn, Johnston, {et~al.}}]{Sharma11}
Sharma, S., Bland-Hawthorn, J., Johnston, K.~V., {et~al.} 2011, ApJ, 730, 20 pp, \dodoi{10.1088/0004-637X/730/1/3}

\bibitem[{Sharma {et~al.}(2016)Sharma, Stello, Bland-Hawthorn, {et~al.}}]{Sharma16}
Sharma, S., Stello, D., Bland-Hawthorn, J., {et~al.} 2016, ApJ, 822, 15 pp, \dodoi{10.3847/0004-637X/822/1/15}

\bibitem[{Silva-Aguirre {et~al.}(2015)Silva-Aguirre, Davies, Basu, {et~al.}}]{Silva-Aguirre15}
Silva-Aguirre, V., Davies, G., Basu, S., {et~al.} 2015, MNRAS, 452, 2127, \dodoi{10.1093/mnras/stv1388}

\bibitem[{Skrutskie {et~al.}(2006)Skrutskie, Cutris, Stiening, {et~al.}}]{MASS06}
Skrutskie, M.~F., Cutris, R.~M., Stiening, R., {et~al.} 2006, The Astronomical Journal, 131, 1163, \dodoi{10.1086/498708}

\bibitem[{Soszyński {et~al.}(2013)Soszyński, Udalski, Szymański, {et~al.}}]{Soszynski13}
Soszyński, I., Udalski, A., Szymański, M.~K., {et~al.} 2013, Acta Astronomica, 63, 21, \dodoi{10.48550/arXiv.1304.2787}

\bibitem[{Spergel {et~al.}(2015)Spergel, Gehrels, Baltay, {et~al.}}]{Spergel15}
Spergel, D., Gehrels, N., Baltay, C., {et~al.} 2015, eprint arXiv, \dodoi{10.48550/arXiv.1503.03757}

\bibitem[{{Sreenivas} {et~al.}(2025){Sreenivas}, {Bedding}, {Huber}, {Crawford}, {Stello}, {Pedersen}, {Li}, \& {Hey}}]{Sreenivas25}
{Sreenivas}, K.~R., {Bedding}, T.~R., {Huber}, D., {et~al.} 2025, arXiv e-prints, arXiv:2502.01899, \dodoi{10.48550/arXiv.2502.01899}

\bibitem[{{Stanek} {et~al.}(1997){Stanek}, {Udalski}, {Szyma{\'N}ski}, {Ka{\L}u{\.Z}ny}, {Kubiak}, {Mateo}, \& {Krzemi{\'N}ski}}]{1997ApJ...477..163S}
{Stanek}, K.~Z., {Udalski}, A., {Szyma{\'N}ski}, M., {et~al.} 1997, \apj, 477, 163, \dodoi{10.1086/303702}

\bibitem[{Stello {et~al.}(2008)Stello, Bruntt, Preston, \& Buzasi}]{Stello08}
Stello, D., Bruntt, H., Preston, H., \& Buzasi, D. 2008, ApJ Letters, 674, L53, \dodoi{10.1086/528936}

\bibitem[{Stello {et~al.}(2013)Stello, Huber, Bedding, {et~al.}}]{Stello13}
Stello, D., Huber, D., Bedding, T., {et~al.} 2013, ApJ, 765, 5, \dodoi{10.1088/2041-8205/765/2/L41}

\bibitem[{Stello {et~al.}(2022)Stello, Saunders, Grunblatt, {et~al.}}]{Stello22}
Stello, D., Saunders, N., Grunblatt, S., {et~al.} 2022, MNRAS, 512, 1677, \dodoi{10.1093/mnras/stac414}

\bibitem[{{Stello} {et~al.}(2017){Stello}, {Zinn}, {Elsworth}, {Garcia}, {Kallinger}, {Mathur}, {Mosser}, {Sharma}, {Chaplin}, {Davies}, {Huber}, {Jones}, {Miglio}, \& {Silva Aguirre}}]{Stello17}
{Stello}, D., {Zinn}, J., {Elsworth}, Y., {et~al.} 2017, \apj, 835, 83, \dodoi{10.3847/1538-4357/835/1/83}

\bibitem[{{STScI}(2011)}]{10.17909/t9059r}
{STScI}. 2011, Kepler/KIC,  STScI/MAST, \dodoi{10.17909/T9059R}

\bibitem[{Tayar {et~al.}(2022)Tayar, Clator, Huber, \& van Saders}]{Tayar22}
Tayar, J., Clator, Z.~R., Huber, D., \& van Saders, J. 2022, ApJ, 927, 11, \dodoi{10.3847/1538-4357/ac4bbc}

\bibitem[{Terry {et~al.}(2023)Terry, Hosek, Lu, {et~al.}}]{Terry23}
Terry, S., Hosek, M., Lu, J., {et~al.} 2023, eprint arXiv, \dodoi{10.48550/arXiv.2306.12485}

\bibitem[{Ting \& Rix(2019)}]{TingRix19}
Ting, Y.-S., \& Rix, H.-W. 2019, ApJ, 878, 16, \dodoi{10.3847/1538-4357/ab1ea5}

\bibitem[{Ulrich(1986)}]{Ulrich86}
Ulrich, R. 1986, ApJ Letters, 306, L37, \dodoi{10.1086/184700}

\bibitem[{Virtanen {et~al.}(2020)Virtanen, Gommers, Oliphant, Haberland, Reddy, Cournapeau, Burovski, Peterson, Weckesser, Bright, {van der Walt}, Brett, Wilson, Millman, Mayorov, Nelson, Jones, Bezanson, van~den Berg, Ivanov, Brumback, Fitzgerald, Gilbert, Gomersall, Redmond, O'Leary, {Hagen}, Mason, Schep, Villa, Turrell, Manohar, Carr, \& Contributors}]{SciPy20}
Virtanen, P.~T., Gommers, R., Oliphant, T.~E., {et~al.} 2020, Nature Methods, 17, 261, \dodoi{10.1038/s41592-019-0686-2}

\bibitem[{Waskom(2021)}]{Waskom21}
Waskom, M.~L. 2021, Journal of Open Source Software, 6, 3021, \dodoi{10.21105/joss.03021}

\bibitem[{White {et~al.}(2011)White, Bedding, Stello, {et~al.}}]{White11}
White, T., Bedding, T., Stello, D., {et~al.} 2011, ApJ, 743, 13, \dodoi{10.1088/0004-637X/743/2/161}

\bibitem[{Wilson {et~al.}(2023)Wilson, Barclay, Powell, {et~al.}}]{Wislon23}
Wilson, R., Barclay, T., Powell, B., {et~al.} 2023, ApJ Supplement Series, 269, 37, \dodoi{10.3847/1538-4365/acf3df}

\bibitem[{Yu {et~al.}(2018)Yu, Huber, Bedding, {et~al.}}]{Yu18}
Yu, J., Huber, D., Bedding, T.~R., {et~al.} 2018, MNRAS, 480, L48, \dodoi{10.1093/mnrasl/sly123}

\end{thebibliography}

\end{document}